\documentclass[prd,aps,showpacs,floats,floatfix,amsfonts,amssymb,amsmath,nofootinbib,superscriptaddress]{revtex4}
\usepackage{epsfig}

\newcommand{\be}{\begin{equation}}\newcommand{\ee}{\end{equation}}
\newcommand{\bea}{\begin{eqnarray}}\newcommand{\eea}{\end{eqnarray}}
\newcommand{\brr}{\begin{array}}\newcommand{\err}{\end{array}}
\newcommand{\bit}{\begin{itemize}}\newcommand{\eit}{\end{itemize}}
\newcommand{\ben}{\begin{enumerate}}\newcommand{\een}{\end{enumerate}}

\newcommand{\ba}{\begin{array}}
\newcommand{\ea}{\end{array}}
\newcommand{\ul}{\underline}
\def\lan{\langle}
\def\lf{\left}

\def\ran{\rangle}

\def\ri{\right}

\def\de{\delta}
\def\te{\theta}
\def\la{\lambda}

\def\1{{_{1}}}\def\2{{_{2}}}

\def\noHe0{:\;\!\!\;\!\!:H_e(0):\;\!\!\;\!\!:}
\def\noHm0{:\;\!\!\;\!\!:H_\mu(0):\;\!\!\;\!\!:}

\begin{document}

\title{Multipartite entangled states in particle mixing}

\author{M. Blasone} \email{blasone@sa.infn.it}
\affiliation{Dipartimento di Matematica e Informatica,
Universit\`a degli Studi di Salerno, Via Ponte don Melillo,
I-84084 Fisciano (SA), Italy} \affiliation{INFN Sezione di Napoli,
Gruppo collegato di Salerno, Baronissi (SA), Italy}

\author{F. Dell'Anno} \email{dellanno@sa.infn.it}
\affiliation{Dipartimento di Matematica e Informatica,
Universit\`a degli Studi di Salerno, Via Ponte don Melillo,
I-84084 Fisciano (SA), Italy} \affiliation{INFN Sezione di Napoli,
Gruppo collegato di Salerno, Baronissi (SA), Italy}
\affiliation{CNR-INFM Coherentia, Napoli, Italy}

\author{S. De Siena} \email{desiena@sa.infn.it}
\affiliation{Dipartimento di Matematica e Informatica,
Universit\`a degli Studi di Salerno, Via Ponte don Melillo,
I-84084 Fisciano (SA), Italy} \affiliation{INFN Sezione di Napoli,
Gruppo collegato di Salerno, Baronissi (SA), Italy}
\affiliation{CNR-INFM Coherentia, Napoli, Italy}

\author{M. Di Mauro}
\affiliation{INFN Sezione di Napoli,
Gruppo collegato di Salerno, Baronissi (SA), Italy}

\author{F. Illuminati} \thanks{Corresponding author} \email{illuminati@sa.infn.it}
\affiliation{Dipartimento di Matematica e Informatica,
Universit\`a degli Studi di Salerno, Via Ponte don Melillo,
I-84084 Fisciano (SA), Italy} \affiliation{INFN Sezione di Napoli,
Gruppo collegato di Salerno, Baronissi (SA), Italy}
\affiliation{CNR-INFM Coherentia, Napoli, Italy}  \affiliation{ISI
Foundation for Scientific Interchange, Viale Settimio Severo 65,
00173 Torino, Italy}

\vspace{2mm}

\begin{abstract}
In the physics of flavor mixing, the flavor states are
given by superpositions of mass eigenstates. By using the occupation
number to define a multiqubit space, the flavor states can be
interpreted as multipartite mode-entangled states. By exploiting a
suitable global measure of entanglement, based on the entropies
related to all possible bipartitions of the system,
we analyze the correlation properties of such states
in the instances of three- and four-flavor mixing.
Depending on the mixing parameters, and, in particular, on
the values taken by the free phases, responsible for the $CP$-violation,
entanglement concentrates in certain bipartitions.
We quantify in detail the amount and the distribution of entanglement
in the physically relevant cases of flavor mixing in quark and neutrino
systems.
By using the wave packet description for localized particles,
we use the global measure of entanglement,
suitably adapted for the instance of multipartite mixed states,
to analyze the decoherence, induced by the free evolution dynamics,
on the quantum correlations of stationary neutrino beams.
We define a decoherence length as the distance associated with
the vanishing of the coherent interference effects among massive
neutrino states.
We investigate the role of the CP-violating phase in the decoherence
process.
\end{abstract}

\date{March 11, 2008}


\maketitle

\section{Introduction}

Quantum entanglement as a physical resource plays a central role
in quantum information and communication science \cite{Nielsen}.
As such, it has been mainly investigated in systems of condensed matter,
atomic physics, and quantum optics.
In fact, such systems offer the most promising possibilities of practical
realizations and implementations of quantum information
tasks. In the domain of particle physics, entanglement has been
discussed mainly in relation to two-body decay, annihilation, and
creation processes, see for instance Refs.
\cite{LeeYang,Inglis,Day,Lipkin,Zralek:1998rp,Bertlmann:2004yg,Cinesi}.
In particular, attention has been focused on the entangled $K_{0}\bar{K}_{0}$ and
$B_{0}\bar{B}_{0}$ states, produced in $e^{+}e^{-}$ collisions
\cite{Selleri,Bertlmann}.
Recently, the entanglement of neutrino pairs, produced in
the tau lepton decay process
$\tau \rightarrow \nu_{\tau}+\nu_{\mu}+e^{-}+\bar{\nu}_{\mu}+\bar{\nu}_{e}$,
has been analyzed in connection with the violation of Bell inequalities \cite{Cinesi}.
\\
A fundamental phenomenon of elementary particles is that of particle
mixing which appears in several instances: quarks, neutrinos, and
the neutral $K$-meson system \cite{Cheng-Li,ParticleData}. Particle
mixing, consisting in a mismatch between flavor and mass,
is at the basis of important effects as neutrino oscillations
and $CP$ violation \cite{Pontecorvo}.
Flavor mixing for the case of three generations is described by the Cabibbo-Kobayashi-Maskawa
(CKM) matrix in the quark instance \cite{Cabibbo,KM}, and by the
Maki-Nakagawa-Sakata-Pontecorvo (MNSP) in the lepton instance \cite{MNSP,Pont}.
The matrix elements represent the transition probabilities from one
lepton (quark) to another. For example, the neutrino mixing is
described by the following relation:
\begin{equation}
\left(\begin{array}{c}
  |\nu_{e}\rangle \\
  |\nu_{\mu}\rangle \\
  |\nu_{\tau}\rangle
\end{array} \right) = \left( \begin{array}{ccc}
  U_{e 1} & U_{e 2} & U_{e 3} \\
  U_{\mu 1} & U_{\mu 2} & U_{\mu 3} \\
  U_{\tau 1} & U_{\tau 2} & U_{\tau 3}
\end{array} \right) \left( \begin{array}{c}
  |\nu_{1}\rangle \\
  |\nu_{2}\rangle \\
  |\nu_{3}\rangle
\end{array}\right) \,,
\label{neutrinomixing}
\end{equation}
where the states $|\nu_{\alpha}\rangle$ with $\alpha=e,\mu,\tau$
denote the neutrino flavor states, the states $|\nu_{i}\rangle$ with
$i=1,2,3$ denote the neutrino mass eigenstates (with masses
$m_{i}$), and $U_{\alpha, i}$ denote the probability amplitudes of
transition of the MNSP matrix $U^{(MNSP)}$. Analogously, for the
quark mixing the CKM matrix connects the weak interaction
eigenstates $(|d'\rangle,|s'\rangle,|b'\rangle)^{T}$ with the strong
interaction eigenstates of the quarks
$(|d\rangle,|s\rangle,|b\rangle)^{T}$; similarly to
Eq.~(\ref{neutrinomixing}), it results
$(|d'\rangle,|s'\rangle,|b'\rangle)^{T} = U^{(CKM)}
(|d\rangle,|s\rangle,|b\rangle)^{T}$. From
Eq.~(\ref{neutrinomixing}), we see that each flavor state is given
by a superposition of mass eigenstates, i.e. $|\nu_{\alpha} \rangle
= U_{\alpha 1} |\nu_{1}\rangle + U_{\alpha 2} |\nu_{2}\rangle +
U_{\alpha 3} |\nu_{3}\rangle$. Let us recall that both
$\{|\nu_{\alpha}\rangle\}$ and $\{|\nu_{i}\rangle\}$ are
orthonormal, i.e. $\langle \nu_{\alpha}|\nu_{\beta}\rangle =
\delta_{\alpha,\beta}$ and $\langle \nu_{i}|\nu_{j}\rangle =
\delta_{i,j}$. Therefore, one can interpret the label $i$ as
denoting a quantum mode, and can legitimately establish the
following correspondence with three-qubit states: $|\nu_{1}\rangle
\equiv |1\rangle_{1} |0\rangle_{2} |0\rangle_{3} \equiv
|100\rangle$, $|\nu_{2}\rangle \equiv |0\rangle_{1} |1\rangle_{2}
|0\rangle_{3} \equiv |010\rangle$, $|\nu_{3}\rangle \equiv
|0\rangle_{1} |0\rangle_{2} |1\rangle_{3} \equiv |001\rangle$, where
$|\rangle_{i}$ denotes states in the Hilbert space for neutrinos
with mass $m_i$. Then, the occupation number allows to interpret the
flavor states as constituted by entangled superpositions of the mass
eigenstates. Quantum entanglement emerges as a direct consequence of
the superposition principle. Let us remark that the Fock space
associated with the neutrino mass eigenstates is physically well
defined. Indeed, at least in principle, the mass eigenstates can be
produced or detected in experiments performing extremely precise
kinematical measurements. For instance, as pointed out by Kayser in
Ref.~\cite{Kayser}, in the process of pion decay $\pi^{+}
\rightarrow \mu^{+} + \nu_{\mu}$, highly precise measurements of the
momenta of the pion and muon will determine the mass squared of the
neutrino $m_{\nu}^{2}$ with an error $\Delta m_{\nu}^{2}$ less than
the mass difference $|m_{i}^{2}-m_{j}^{2}| \quad (i\neq j =1,2,3)$.
Thus, the ``physical neutrino'' $|\nu_{i}\rangle$ involved in each
event of the process is fully determined \cite{Kayser}. This kind of
experiment will lead
to the destruction of the oscillation phenomenon. \\
Therefore, entanglement is established among field modes,
although the quantum mechanical state is a single-particle one.
This is in complete analogy to the mode entanglement defined for single-photon
states of the radiation field or the mode entanglement introduced
for systems of identical particles \cite{Zanardi}: In all these
instances, entanglement is established not between particles, but
rather between field modes. In the particle physics
instance, the multipartite flavor states can be seen as a generalized
class of $W$ states. The latter are multipartite entangled states that
occur in a variety of diverse physical systems and can be
engineered even with pure quantum optical elements \cite{MultiphotonReport}.
From a theoretical standpoint, the concept of mode entanglement in single-particle
states has been widely discussed in the literature and is by now well established
\cite{singpart1,singpart2,singpart3,Zanardi}, and linear
optical scheme has been proposed to demonstrate multipartite entanglement
of single-photon $W$ states \cite{singpart4}. Experimental realizations
include the teleportation of a single-particle entangled qubit \cite{singpart5},
the quantum state reconstruction of single-photon entangled Fock states \cite{singpart6},
and the homodyne tomography characterization of dual-mode optical qubits using a single
photon delocalized over two optical modes \cite{singpart7}.
Among the experimental proposals, we should mention a scheme for quantum cryptography
using single-particle entanglement \cite{singpart8}.
Moreover, remarkably, the nonlocality of single-photon states has been experimentally
demonstrated by double homodyne measurements \cite{singpartexp}, thus verifying a
long-standing theoretical prediction \cite{Tan,Hardy}. Very recently, the existing
schemes to probe nonlocality in single-particle states have been generalized to
include massive particles of arbitrary type \cite{Vedral}, thus paving the way to
the study of single-particle entanglement in a variety of diverse systems including
atoms, molecules, nuclei, and elementary particles.

Concerning the neutrino system, the main difference between the
single-photon states and the single-particle neutrino states is
related to the spatial separability of modes. For instance, the
polarization modes of polarization-entangled single-photon states
can be easily spatially separated by means of a polarizing beam
splitter. On the contrary, at present, it is not available a beam
splitter analog for neutrinos. However, it is worth noting to recall
that spatial separability and nonlocality are not necessary
requirements for entanglement \cite{singpart3}. Nevertheless, the
spatial separation between massive neutrino states emerges in the
dynamics of the free evolution in the wave packet approach
\cite{Nussinov,GiuntiKim,Giunti2,Giunti:2008cf}. In quantum theory
localized particles are described by wave packets; moreover, during
the free propagation, the different mass eigenstates
$|\nu_{i}\rangle$ in the packet travel at different speeds. Thus,
the evolution leads to a spatial separation along the propagation
direction (time delay) of the mass eigenstates $|\nu_{i}\rangle$,
and the difference between their arrival times at a given detector
is observable \cite{Beuthe,Giunti:2007ry}. The ``decoherence''
induced by the free evolution leads to a degradation and even to a
complete destruction of the oscillation phenomenon
\cite{GiuntiKim,Giunti2,Giunti:2008cf}. It is interesting to
investigate the influence of the decoherence on the quantum
correlation of the multipartite entangled mass eigenstates.

The issue of mode entanglement in single-particle states of elementary particle
physics has been recently addressed by the study of the dynamical behavior of
bipartite and multipartite flavor entanglement in neutrino oscillation
\cite{neutrinooscillatentang}. In the present paper we characterize the correlation
properties of the multipartite single-particle states that emerge in the context of
lepton or quark mixing. These states turn out to be generalized $W$-like entangled states.
By resorting to a suitable measure of global entanglement, we analyze in detail
their properties for different occurrences of flavor mixing and particle oscillations
both in the quark and in the leptonic sectors.
Furthermore, by using the wave packet description for the free-propagating neutrino states,
we analyze the dynamical behavior of the multipartite entanglement in the
phenomenon of neutrino oscillations.
The paper is organized as follows:
In Section \ref{MultipEnt} we discuss the main aspects of different measures of
multipartite entanglement. Following the approach of Ref. \cite{Oliveira}, we introduce
a characterization of multipartite quantum correlations based on suitable entanglement
measures for all the possible bipartitions of the $N$-partite system.
In Section \ref{flavmixWstates} we recall the formalism of flavor mixing in order
to define generalized classes of three-partite $W^{(3)}$ and four-partite $W^{(4)}$ states.
In Section \ref{entWN} we study in detail the behavior of entanglement for
the $W^{(3)}$ and $W^{(4)}$ states as a function of the free phases in the case
of maximal mixing. In Section \ref{QuarksandNeutrino}, we apply the general formalism developed
in the previous Sections to the quantification of multipartite entanglement
in the most relevant cases of quark and neutrino flavor mixing.
Finally, in Section \ref{DecoStatBeam} by using the wave packet treatment,
we analyze the effect of propagation-induced decoherence on multipartite entanglement.

\section{Multipartite entanglement}
\label{MultipEnt}

In this Section we will briefly discuss the problem of quantifying
multipartite entanglement in relation to global aspects and statistical
properties, and introduce measures particularly suitable for
our purposes. For recent, detailed reviews on the qualification,
quantification, and applications of entanglement, see Refs.
\cite{EntRevHorodecki,EntRevFazio,EntRevPlenio}. Concerning bipartite pure
states, entanglement is very well characterized by proper and efficient
measures. In fact, for a bipartite pure state $\rho_{12}$ the von
Neumann entropy $E_{vN} \,=\, -Tr_{1}[\rho_{1}\log_{2}\rho_{1}]$,
for the reduced density matrix $\rho_{1}=Tr_{2}[\rho_{12}]$,
completely determines the amount of entanglement
\cite{Popescu}. For a given bipartition, $E_{vN}$ has its
maximum $\log_{2}d$, where $d$ denotes the minimum between the
dimensions of the two parties. For bipartite mixed states, several
entanglement measures have been proposed
\cite{EntFormDistill,EntRelEntr,Negativity}. Although providing
interesting operative definitions, the entanglement of formation
and of distillation \cite{EntFormDistill} are very hard to
compute. A celebrated result is the Wootters formula for the
entanglement of formation for two-qubit mixed states
\cite{Wootters}. An alternative measure, closely related to the
entanglement of formation, is the concurrence (the entanglement of
formation is a monotonically increasing function of the
concurrence) \cite{CoffKundWoot}. The same difficulties of
computation are encountered with the relative entropy of
entanglement \cite{EntRelEntr}.
At present a computable entanglement monotone is the logarithmic negativity
$E_{\mathcal{N}}$, based on the requirement of
positivity of the density operator under partial transposition
$E_{\mathcal{N}} = \log_{2} \parallel \tilde{\rho}_{12} \parallel_{1}$,
where $\parallel \cdot \parallel_{1}$ denotes the trace norm, i.e.
$\parallel \mathcal{O} \parallel_{1} = Tr[\sqrt{\mathcal{O}^{\dag}\mathcal{O}}]$
for any Hermitian operator $\mathcal{O}$ \cite{Negativity}.
The so-called {\it bona fide} density matrix $\tilde{\rho}_{12}$
is obtained through the partial transposition with respect to one mode,
say mode $2$, of $\rho_{12}$, i.e. $\tilde{\rho}_{12} \equiv \rho_{12}^{PT \, 2}$.
Given an arbitrary orthonormal product basis $|i_{1}\,, j_{2}\rangle$,
the matrix elements of $\tilde{\rho}_{12}$ are determined by the relation
$\langle i_{1}\,, j_{2}| \tilde{\rho}_{12} | k_{1}\,, l_{2}\rangle =
\langle i_{1}\,, l_{2}| \rho_{12} | k_{1}\,, j_{2}\rangle$.
Obviously, for pure states such a measure provides the
same results as the von Neumann entropy. \\
The challenge of quantifying entanglement becomes much harder
in multipartite systems. Important achievements have been reached in
understanding the different ways in which multipartite systems can be entangled.
The intrinsic nonlocal character of entanglement imposes its invariance
under any local quantum operations; therefore, equivalence classes
of entangled states can be defined through the group of reversible stochastic local
quantum operations assisted by classical communication (SLOCCs) \cite{SLOCC}.
Such an approach allows to demonstrate that three and four qubits can be entangled,
respectively, in two and nine inequivalent ways \cite{2diffwayent,9diffwayent}.
In particular, all three-qubit entangled states are related to two fundamental
classes of states: the $GHZ$ state $|GHZ^{(3)}\rangle$
and the $W$ state $|W^{(3)}\rangle$ \cite{2diffwayent,GHZst}.
In the $N$-partite instance, such states are defined as:
\begin{eqnarray}
&&|GHZ^{(N)}\rangle \,=\, \frac{1}{\sqrt{2}} (|000\ldots 0\rangle +
|111\ldots 1\rangle) \,,
\label{GHZN} \\
&& \nonumber \\
&&|W^{(N)}\rangle \,=\, \frac{1}{\sqrt{N}}
(|100\ldots 0\rangle + |010\ldots 0\rangle + |001\ldots 0\rangle +
\ldots |000\ldots 1\rangle) \,.
\label{WN}
\end{eqnarray}
The $GHZ$ and $W$ states are fully symmetric, i.e. invariant under the exchange of any two
qubits, and greatly differ each other in their correlations properties. The $GHZ$ state
possesses maximal $N$-partite entanglement, i.e. it violates Bell inequality maximally;
on the other hand, it lacks bipartite entanglement. For instance, in the case $N=3$,
abandoning one mode, the resulting mixed two-mode state turns out to be separable.
The $W$ states possess less $N$-partite entanglement, but maximal $K$-partite entanglement
($K < N$) in the $K$-reduced states.

Several attempts have been done to introduce efficient
entanglement measures for multipartite systems.
The characterization of the quantum correlations
through a measure embodying a collective property of the system,
should be based on the introduction of quantities invariant under
local transformations.
A successful step in this direction has been put forward by
Coffman, Kundu, and Wootters. Studying the distributed
entanglement in systems of three qubits, they defined the
so-called residual, genuine tripartite entanglement, or $3$-tangle,
a quantity constructed in terms of the squared concurrences
associated with the global three qubit state and the reduced
two-qubit states \cite{CoffKundWoot}.
While successfully detecting the genuine tripartite entanglement
in the state $|GHZ^{(3)}\ran$, the $3$-tangle vanishes
if computed for the state $|W^{(3)}\ran$, thus being not appropriate
for this class of states.
Several generalizations of the $3$-tangle have been proposed
\cite{MP3tangle}.
The Schmidt measure, defined as the minimum of $\log_{2} r$ with $r$ being the minimum
of the number of terms in an expansion of the state in product basis,
has been proposed by Eisert and Briegel \cite{EisertBriegel}.
The measure vanishes if and only if the state is fully product,
thus it does not discriminate between genuine multipartite and bipartite entanglement.
However, the Schmidt measure is able to distinguish the $GHZ$ and the $W$ states;
for instance, it yields the value $1$ for $|GHZ^{(N)}\ran$
and the values $\log_{2}N$ for the $|W^{(N)}\ran$ state
(considering $N$-partitions of the system).
Multipartite entanglement can be characterized also by the distance
of the entangled state to the nearest separable state;
this is the geometric measure \cite{geometricmeasure}.
Simpler proposals are given in terms of functions of bipartite entanglement
measures \cite{Wallach,Brennen,Scott,Oliveira,Pascazio}.
An example of this kind of proposals is the global entanglement of
Meyer and Wallach, that is defined as the sum of concurrences between
one qubit and all others \cite{Wallach}, and can be expressed as the
average subsystem linear entropy \cite{Brennen}.
A generalization of the global entanglement has been introduced
by Rigolin {\it et al.}, using the set of the mean linear entropies
of all possible bipartitions of the whole system \cite{Oliveira}.
Recently, another approach has been proposed, based on the distribution
of the purity of a subsystem over all possible bipartitions of the total system
\cite{Pascazio}.

\subsection{Average von Neumann entropy}
\label{avervonNeum}

We intend to analyze the entanglement properties of a generalized
class of $W$ states (finite-dimensional pure states). To this aim,
we adopt an approach similar to that of Refs.
\cite{Wallach,Brennen,Scott,Oliveira,Pascazio}, thus
characterizing the entanglement through measures defined on the possible
bipartitions of the system. As we are dealing with pure states, we define
as proper measure of multipartite entanglement a functional of the
von Neumann entropy averaged on a given bipartition of the system.
Let $\rho=|\psi\ran\lan \psi|$ be the density operator
corresponding to a pure state $|\psi\ran$, describing the system
$S$ partitioned into $N$ parties. Let us consider the bipartition
of the $N$-partite system $S=\{S_{1},S_{2},\ldots,S_{N}\}$ in two
subsystems $S_{A_{n}}=\{S_{i_{q}},S_{i_{2}},\ldots,S_{i_{n}}\}$,
with $1\leq i_{1}<i_{2}<\ldots <i_{n}\leq N$ $(1\leq n <N)$, and
$S_{B_{N-n}}=\{S_{j_{1}},S_{j_{2}},\ldots,S_{j_{N-n}}\}$, with
$1\leq j_{1}<j_{2}<\ldots <j_{N-n} \leq N$, and $i_{q}\neq j_{p}$.
We denote by
\begin{equation}
\rho_{A_{n}} \equiv \rho_{i_{1},i_{2},\ldots,i_{n}} \,=\,
Tr_{B_{N-n}}[\rho] \,=\, Tr_{j_{1},j_{2},\ldots,j_{N-n}}[\rho] \,
\label{reducedrho}
\end{equation}
the density matrix reduced with respect to the subsystem
$S_{B_{N-n}}$. The von Neumann entropy associated with such a
bipartition will be given by
\begin{equation}
E_{vN}^{(A_{n};B_{N-n})} \,=\,
-Tr_{A_{n}}[\rho_{A_{n}}\log_{2}\rho_{A_{n}}] \,.
\label{vonNeumAn}
\end{equation}
At last, we define the average von Neumann entropy
\begin{equation}
\langle E_{vN}^{(n:N-n)} \rangle \,=\, \left(%
\begin{array}{c}
  N \\
  n \\
\end{array}%
\right)^{-1} \; \sum_{A_{n}} E_{vN}^{(A_{n};B_{N-n})} \,,
\label{avnvonNeum}
\end{equation}
where the sum is intended over all the possible bipartitions of
the system in two subsystems each with $n$ and $N-n$ elements
$(1\leq n <N)$. \\
For instance, in the simple cases of a three qubit
states, as the states $\rho_{W^{(3)}} \,=\, |W^{(3)}\ran\lan
W^{(3)}|$ and $\rho_{GHZ^{(3)}} \,=\, |GHZ^{(3)}\ran\lan
GHZ^{(3)}|$, only unbalanced bipartitions of two subsystems can be
considered. Straightforward calculations yield
\begin{eqnarray}
&& E_{21}^{(3)} \,\equiv\, E_{vN}^{(A_{2};B_{1})}(\rho_{W^{(3)}})
\,=\, \langle E_{vN}^{(2:1)}(\rho_{W^{(3)}}) \rangle \,=\,
\log_{2}3 -\frac{2}{3} \,\simeq\, 0.918296 \,, \label{E21W3}
\\
&&E_{vN}^{(A_{2};B_{1})}(\rho_{GHZ^{(3)}}) \,=\, \langle
E_{vN}^{(2:1)}(\rho_{GHZ^{(3)}}) \rangle \,=\, 1 \,.
\end{eqnarray}
On the other hand, for a four-qubit state we have both unbalanced, i.e.
$(S_{A_{3}},S_{B_{1}})$, and balanced bipartitions, i.e.
$(S_{A_{2}},S_{B_{2}})$. For the state $\rho_{W^{(4)}} \,=\,
|W^{(4)}\ran\lan W^{(4)}|$, we get
\begin{eqnarray}
&& E_{31}^{(4)} \,\equiv\, E_{vN}^{(A_{3};B_{1})}(\rho_{W^{(4)}})
\,=\, \langle E_{vN}^{(3:1)}(\rho_{W^{(4)}}) \rangle \,=\,
2-\frac{3}{4}\log_{2}3 \,\simeq\, 0.811278 \,, \label{E31W4}
\\
&& E_{22}^{(4)} \,\equiv\, E_{vN}^{(A_{2};B_{2})}(\rho_{W^{(4)}})
\,=\, \langle E_{vN}^{(2:2)}(\rho_{W^{(4)}}) \rangle \,=\, 1  \,.
\label{E22W4}
\end{eqnarray}
Of course, all the measures evaluated on the state $\rho_{GHZ^{(4)}}$
give the maximal, normalized value $1$. It is worth noting
that in order to characterize the multipartite entanglement in a $N$-partite
system, the number of bipartite measures grows with $N$.

\subsection{Average logarithmic negativity}
\label{averlogNegativity}

As well known, the entropic measures cannot be used to quantify the entanglement of
mixed states. In order to measure the multipartite entanglement of
mixed states, and following the same procedure as in the previous subsection,
we introduce a generalized version of the logarithmic
negativity \cite{Negativity}.
Let $\rho$ be a multipartite mixed state associated with a system $S$,
partitioned into $N$ parties. Again we consider the bipartition of the
$N$-partite system $S$ into two subsystems $S_{A_{n}}$ and $S_{B_{N-n}}$.
We denote by
\begin{equation}
\tilde{\rho}_{A_{n}} \equiv  \rho^{PT\, B_{N-n} }
\,=\,  \rho^{PT\, j_{1},j_{2},\ldots,j_{N-n} }
\label{rhoPT}
\end{equation}
the {\it bona fide} density matrix, obtained by the partial transposition of $\rho$
with respect to the parties belonging to the subsystem $S_{B_{N-n}}$.
The logarithmic negativity associated with the fixed bipartition will be given by
\begin{equation}
E_{\mathcal{N}}^{(A_{n};B_{N-n})} \,=\,
\log_{2} \parallel \rho_{A_{n}}\log_{2} \tilde{\rho}_{A_{n}} \parallel_{1} \,.
\label{lognegAn}
\end{equation}
Finally, we define the average logarithmic negativity
\begin{equation}
\langle E_{\mathcal{N}}^{(n:N-n)} \rangle \,=\, \left(%
\begin{array}{c}
  N \\
  n \\
\end{array}%
\right)^{-1} \; \sum_{A_{n}} E_{\mathcal{N}}^{(A_{n};B_{N-n})} \,,
\label{avnlogneg}
\end{equation}
where the sum is intended over all the possible bipartitions of the system.

\section{Generalized $W$ states in flavor mixing}
\label{flavmixWstates}

In this Section, we consider generalized $W$ states of the form:
\begin{equation}
|W^{(N)}(\alpha_{1},\alpha_{2},\ldots,\alpha_{N})\ran \,=\,
\sum_{k=1}^{N} \alpha_{k} \,
|\delta_{1,k},\delta_{2,k},\ldots,\delta_{N,k}\ran \equiv
\sum_{k=1}^{N} \alpha_{k} \, |\nu_{k}^{(N)}\ran \,, \qquad
\sum_{k=1}^{N} |\alpha_{k}|^{2} \,=\, 1 \,,
\label{WNdef}
\end{equation}
where $\delta_{m,n}$ denotes the Kronecker delta. In particular,
we will consider the cases corresponding to $N=3,4$. Moreover, we
will adopt a parametrization for $\{\alpha_{k}\}$ commonly used in
the domain of elementary particle physics, and associated with the
phenomena of $N$-flavor mixing, i.e. quark and neutrino mixing \cite{Cheng-Li}. \\
The orthonormal set of flavor states $|\psi_{l}^{(N)}\ran$ are defined through the
application of the $N\times N$ mixing matrix $U^{(Nf)}$ to the
basis vectors $|\nu_{k}^{(N)}\ran$, i.e. $|\psi_{l}^{(N)}\ran
\,=\, \sum_{k=1}^{N} U_{l,k}^{(Nf)} |\nu_{k}^{(N)}\ran$
$(l,k=1,\ldots,N)$. An $N \times N$ unitary matrix contains, in
general, $ N^2$ independent parameters. Each of the $2N$ fields
(two for each lepton generation) can absorb one phase. Moreover,
there is an unobservable overall phase, so we are left with
$(N-1)^2$ independent real parameters. Among these, $\frac{N(N-1)}{2}$
are rotation angles, or mixing angles, and the remaining
$\frac{(N-1)(N-2)}{2}$ are phases, which are responsible for $CP$
violation. Applying this formalism, we determine $N$ orthonormal flavor
states $|\psi_{l}^{(N)}\ran$, that belong to the class of generalized $W$
states defined by Eq.~(\ref{WNdef}).

\subsection{Generalized W$^{(3)}$ states from three-flavor mixing matrix}
\label{W3states}

In the case of mixing among three generations (either leptons
or quarks), the standard parametrization of a $3\times 3$
unitary mixing matrix is given by \cite{Cheng-Li}:
\bea \label{fermix3} |\underline{\nu}_f\ran &=&
U(\tilde{\theta},\delta) \, |\underline{\nu}_m\ran
\\ [2mm]
\label{CKMmatrix} U(\tilde{\theta},\delta)&=&\lf(\ba{ccc}
c_{12}c_{13} & s_{12}c_{13} & s_{13}e^{-i\de} \\
-s_{12}c_{23}-c_{12}s_{23}s_{13}e^{i\de} &
c_{12}c_{23}-s_{12}s_{23}s_{13}e^{i\de} & s_{23}c_{13} \\
s_{12}s_{23}-c_{12}c_{23}s_{13}e^{i\de} &
-c_{12}s_{23}-s_{12}c_{23}s_{13}e^{i\de} & c_{23}c_{13} \ea\ri)\,
\,,
\eea
where $|\underline{\nu}_f\ran \,=\, \left( |\nu_e\ran,|\nu_\mu \ran,
|\nu_\tau \ran \right)^{T}$ are the states with definite flavor and
$|\underline{\nu}_m\ran \,=\, \left( |\nu_1\ran,|\nu_2\ran,
|\nu_3\ran \right)^{T}$ those with definite masses. In
Eqs.~(\ref{fermix3}) and (\ref{CKMmatrix}), the following
shorthand notation has been adopted:
$(\tilde{\theta},\delta)\equiv(\theta_{12},\theta_{13},\theta_{23};\delta)$,
$c_{ij}\equiv\cos\theta_{ij}$ and $s_{ij}\equiv\sin\theta_{ij}$.
In this case, we have three mixing angles $\theta_{12}$,
$\theta_{13}$, $\theta_{23}$, and a free phase $\delta$. It can be
shown that the values of these parameters for which the three
flavor mixing is maximal are \cite{KM}:
\bea
\label{maxparam3} \te_{12}^{max}= \frac{\pi}{4}; \qquad
\te_{23}^{max}= \frac{\pi}{4}; \qquad \te_{13}^{max}=
\arccos{\sqrt{\frac{2}{3}}}; \qquad \de^{max} =\frac{\pi}{2} \,
.\eea
In correspondence of these values, the matrix elements in
Eq.~(\ref{CKMmatrix}) have all the same modulus
$\frac{1}{\sqrt{3}}$. \\
For $N=3$, we define the generalized class of three-qubit $W$ states
as those generated by means of the following matrix, which is
obtained by the above mixing matrix upon multiplication of the
third column by $e^{i\de}$:
\bea \label{fermix3b}
|\underline{W}^{(3)}(\tilde{\theta};\delta)\ran &\equiv& U^{(3
f)}(\tilde{\theta},\delta) \, |\underline{\nu}^{(3)}\ran
\\ [2mm]
\label{CKMmatrixb} U^{(3 f)}(\tilde{\theta},\delta)&=&
U(\tilde{\theta},\delta) \lf(\ba{ccc}
1 & 0 & 0 \\
0 &
1 &0\\
0 & 0&  e^{i\de} \ea\ri)\,, \eea
where $|\underline{W}^{(3)}(\tilde{\theta};\delta)\ran \,=\,
\left(
|W_{e}^{(3)}(\tilde{\theta},\delta)\ran,|W_{\mu}^{(3)}(\tilde{\theta},\delta)\ran,
|W_{\tau}^{(3)}(\tilde{\theta},\delta)\ran \right)^{T}$ and
$|\underline{\nu}^{(3)}\ran \,=\, \left(
|\nu_{1}^{(3)}\ran,|\nu_{2}^{(3)}\ran, |\nu_{3}^{(3)}\ran
\right)^{T}$. The entanglement properties of the
states associated with matrices (\ref{CKMmatrix}) and
(\ref{CKMmatrixb}) are identical. When all the mixing parameters
are chosen to be maximal as in Eq.~(\ref{maxparam3}), the matrix
$U^{(3 f)}$ becomes:
\bea
\label{fermixmax}
U^{(3 f)}_{max}=
\frac{1}{\sqrt{3}}\lf(\ba{ccc}
1 & 1 & 1 \\
i y & i y^2 & i \\
i y^2 &i y & i \ea\ri) \, .
\eea
with $y=\exp{(2 i \pi/3) }$. In the case of maximal mixing,
all the states possess the same entanglement of  $|W^{(3)}\ran$:
\begin{equation}
E_{vN}^{(A_{2};B_{1})}(|\underline{W}^{(3)}(\tilde{\theta}^{max};\delta^{max})\ran)
\,=\, \langle
E_{vN}^{(2:1)}(|\underline{W}^{(3)}(\tilde{\theta}^{max};\delta^{max})\ran)
\rangle \,=\, E_{21}^{(3)} \,, \label{refentW3}
\end{equation}
where $E_{21}^{(3)}$ is defined in Eq.~(\ref{E21W3}). In the
next Section we will analyze the entanglement properties of the
$|W_{\alpha}^{(3)}(\tilde{\theta},\delta)\ran$ states, and their
behavior as a function of the mixing parameters.

\subsection{Generalized W$^{(4)}$ states from four-flavor mixing matrix}
\label{W4states}

Let us now consider the four-flavor mixing $(N=4)$. In particle physics, such a case could
be realized, for instance, by a situation in which there are three active neutrino types
and an extra one, non interacting, the so-called ``sterile'' neutrino. Obviously, such
states correspond to physically realizable situations in optical and condensed matter
systems. The corresponding four-flavor mixing matrix
$U^{(4f)}(\tilde{\theta},\tilde{\delta})$ will be built on 9
independent parameters, 6 mixing angles and 3 phases, i.e.
$(\tilde{\theta};\tilde{\delta})=
(\theta_{12},\theta_{13},\theta_{14},\theta_{23},\theta_{24},\theta_{34};\delta_{14},\delta_{23},\delta_{34})$.
Explicitly, the mixing matrix for four flavors can be written as the following
product of elementary matrices:
\begin{equation}
U^{(4f)}(\tilde{\theta};\tilde{\delta})\,=\,
U_{34}(\theta_{34},\delta_{34})U_{24}(\theta_{24})U_{23}(\theta_{23},\delta_{23})
U_{14}(\theta_{14},\delta_{14})U_{13}(\theta_{13})U_{12}(\theta_{12}) U_\de(\de_{14})\,,
\label{matmix4}
\end{equation}
where
\begin{eqnarray}
U_\de(\de_{14})&=& \lf(\ba{cccc}
1 & 0 &0&0\\
0 & 1 &0&0\\
0&0&1&0\\
0&0&0&e^{i \de_{14}}
 \ea\ri); \quad
U_{12}=\lf(\ba{cccc}
\cos\te_{12} & \sin\te_{12} &0&0\\
-\sin\te_{12} & \cos\te_{12} &0&0\\
0&0&1&0\\
0&0&0&1
 \ea\ri) ;
\quad U_{13}= \lf(\ba{cccc}
\cos\te_{13} &0& \sin\te_{13} &0\\
0&1&0&0\\
-\sin\te_{13}&0 & \cos\te_{13} &0\\
0&0&0&1
 \ea\ri)
\\ [2mm]
U_{14}&=& \lf(\ba{cccc}
\cos\te_{14} &0 &0& e^{-i \de_{14}}\sin\te_{14}\\
0&1&0&0\\
0&0&1&0\\
-e^{i \de_{14}}\sin\te_{14}&0&0 & \cos\te_{14} \ea\ri); \quad
U_{23}= \lf(\ba{cccc}
1&0&0&0\\
0&\cos\te_{23} &  e^{-i \de_{23}}\sin\te_{23} &0\\
0&-e^{i \de_{23}}\sin\te_{23}&\cos\te_{23}&0\\
0&0&0&1
 \ea\ri)
\\ [2mm]
U_{24}&=& \lf(\ba{cccc}
1&0&0&0\\
0&\cos\te_{24}&0 &  \sin\te_{24} \\
0&0&1&0\\
0&-\sin\te_{24}&0&\cos\te_{24}
 \ea\ri);
\quad U_{34}= \lf(\ba{cccc}
1&0&0&0\\
0&1&0&0\\
0&0&\cos\te_{34} & e^{-i \de_{34}} \sin\te_{34} \\
0&0&-e^{i \de_{34}}\sin\te_{34}&\cos\te_{34}
 \ea\ri)
\end{eqnarray}
Analogously to the definition (\ref{fermix3b}) given in subsection \ref{W3states},
the class of generalized four-qubit $W$ states can be defined as
\begin{equation}
|\underline{W}^{(4)}(\tilde{\theta};\tilde{\delta})\ran \,\equiv\,
{ U}^{(4f)}(\tilde{\theta};\tilde{\delta}) \, |\underline{\nu}^{(4)}\ran \,.
\label{fermix4}
\end{equation}
The matrix (\ref{matmix4}) is maximal, i.e. all elements have the
same modulus $1/2$, for the following set of values: \bea
\label{thetamax4} &&\te_{12}^{max}= \te_{34}^{max}=\frac{\pi}{4};
\qquad \te_{14}^{max}= \te_{23}^{max}= \frac{\pi}{6}; \qquad
\te_{13}^{max}= \arccos{\sqrt{\frac{2}{3}}}; \qquad
\te_{24}^{max}= \arcsin{\sqrt{\frac{1}{3}}};
\\ [3mm]
&&\de_{14}^{max} =\phi; \qquad \de_{23}^{max}
=\pi-\phi; \qquad \de_{34}^{max} =\phi\, .
\label{deltamax4}
\eea
For the choices (\ref{thetamax4}) and (\ref{deltamax4}),
$U^{(4f)}_{max}(\phi)$ takes the simple form
\begin{equation}
U^{(4f)}_{max}(\phi) \,=\, \frac{1}{2}
\lf(
\ba{cccc}
1 & 1 & 1 & 1 \\
-1 & 1 & -e^{i\phi} & e^{i\phi} \\
-1 & -1 & 1 & 1\\
1 & -1 & -e^{i\phi} &e^{i\phi}
\ea
\ri).
\label{U4fmax}
\end{equation}
All the states
$|\underline{W}^{(4)}(\tilde{\theta}^{max};\tilde{\delta}^{max})\ran$
exhibit the same amount of entanglement of the standard four-qubit
$|W^{(4)}\ran$ state:
\begin{eqnarray}
E_{vN}^{(A_{3};B_{1})}(|\underline{W}^{(4)}(\tilde{\theta}^{max};\tilde{\delta}^{max})\ran)
&=& \lan
E_{vN}^{(3:1)}(|\underline{W}^{(4)}(\tilde{\theta}^{max};\tilde{\delta}^{max})\ran)
\ran  \,=\, E_{31}^{(4)} \,.
 \\ && \nonumber \\
E_{vN}^{(A_{2};B_{2})}(|\underline{W}^{(4)}(\tilde{\theta}^{max};\tilde{\delta}^{max})\ran)
&=& \lan
E_{vN}^{(2:2)}(|\underline{W}^{(4)}(\tilde{\theta}^{max};\tilde{\delta}^{max})\ran)
\ran \,=\, E_{22}^{(4)} \,,
\end{eqnarray}
for any bipartition $(A_{2}\,,B_{2})$ and $(A_{3}\,,B_{1})$.
$E_{31}^{(4)}$ and $E_{22}^{(4)}$ are given in Eqs.~(\ref{E31W4}) and (\ref{E22W4}), respectively.

\section{The correlation properties of $ |W^{(N)}\ran $ states}
\label{entWN}

In this Section we analyze the correlation properties of the
class of W-like states defined by
Eqs.~(\ref{fermix3b}) and (\ref{fermix4}). Such
properties are completely determined by the free parameters
of the mixing matrix formalism, i.e. the rotation angles
$\theta_{ij}$ and the phases $\de_{ij}$. Let us recall that for
$N=3$ we have three angles and one phase, while for $N=4$ we have
six angles and three phases.
In our formalism, the state associated with the first row of the matrices
$U^{(3 f)}(\tilde{\theta},\delta)$ and $U^{(4 f)}(\tilde{\theta},\tilde{\delta})$,
i.e. the states $|W_e^{(3)}(\tilde{\theta};\delta)\ran$
and $|W_e^{(4)}(\tilde{\theta};\delta)\ran$, respectively,
reduce to standard $3$-qubit and $4$-qubit W states by fixing the rotation angles
to their maximal values  $\theta_{ij}^{max}$,
according to Eqs.~(\ref{maxparam3}) and (\ref{thetamax4}).
Therefore, in the instance of $N$-flavor W states with maximal mixing angles,
i.e. $|\underline{W}^{(N)}(\tilde{\theta}^{max};\tilde{\delta})\ran$,
there exists a subspace (of dimension $N-1$), that is
is orthogonal to the $|W^{(N)}\ran$ state and is
spanned by the vectors
 $\Big\{|W_{\alpha_{2}}^{(N)}(\tilde{\theta}^{max};\tilde{\delta})\ran,
\ldots |W_{\alpha_{N}}^{(N)}(\tilde{\theta}^{max};\tilde{\delta})\ran \Big\}$.
For simplicity, in the following we will restrict ourselves to the study of
the entanglement properties of such a subclass of generalized $|W^{(N)}\ran$ states,
which are parameterized by the phases of the mixing matrix.

\subsection{Case of $|\underline{W}^{(3)}(\tilde{\theta}^{max};\delta)\ran$ states}
\label{entW3}

First, we discuss the entanglement properties of $3$-partite W states
$U^{(3 f)}(\tilde{\theta}^{max},\delta)$; in particular, we study the
dependence of entanglement on the phase $\de$, with the rotation angles
$\theta_{ij}$ at their maximal values $\theta_{ij}^{max}$,
given by Eq.~(\ref{maxparam3}). In this way, we obtain a set of three
orthogonal generalized $W$ states $|W_{\alpha}^{(3)}(\delta)\ran \equiv
|W_{\alpha}^{(3)}(\tilde{\theta}^{max};\delta)\ran$ $(\alpha=e,\mu,\tau)$, of
which the first one is the usual $|W^{(3)}\ran$ state.
Correspondingly, the matrix $U^{(3 f)}$ is specialized to
\bea {U}^{(3 f)}(\delta)&=& \frac{1}{\sqrt{3}}
\lf(\ba{ccc} 1 & 1 & 1 \\ [1.5mm]
-\frac{1}{2}(\sqrt{3} +e^{i\de}) & \frac{1}{2}(\sqrt{3} -e^{i\de}) & e^{i\de} \\ [1.5mm]
\frac{1}{2}(\sqrt{3} -e^{i\de}) & -\frac{1}{2}(\sqrt{3} +e^{i\de})
& e^{i\de} \ea\ri) \,.
\label{CKMmatrixdelta}
\eea
Let us compute the quantities $E_{vN}^{(A_{2};B_{1})}$
and $\lan E_{vN}^{(2:1)} \ran$, as defined by
Eqs.~(\ref{vonNeumAn}) and (\ref{avnvonNeum}) in Section
\ref{avervonNeum}. We get:
\begin{eqnarray}
&&E_{vN\,e}^{(1,2;3)} \,=\,
E_{vN\,e}^{(1,3;2)} \,=\,
E_{vN\,e}^{(2,3;1)} \,=\,
E_{vN\,\mu}^{(1,2;3)} \,=\,
E_{vN\,\tau}^{(1,2;3)} \,=\,
\log_{2}3 -\frac{2}{3} \,,
\label{E3a}
\\ && \nonumber \\
&&E_{vN\,\mu}^{(1,3;2)} \,=\,
E_{vN\,\tau}^{(2,3;1)} \,=\,
-\lf(\frac{1}{3}-\frac{\cos\delta}{2\sqrt{3}} \ri)
\log_{2}\lf[\frac{1}{3}-\frac{\cos\delta}{2\sqrt{3}} \ri]
-\lf(\frac{2}{3}+\frac{\cos\delta}{2\sqrt{3}} \ri)
\log_{2}\lf[\frac{2}{3}+\frac{\cos\delta}{2\sqrt{3}} \ri] \,,
\label{E3b}
\\ && \nonumber \\
&&E_{vN\,\mu}^{(2,3;1)} \,=\,
E_{vN\,\tau}^{(1,3;2)} \,=\,-\lf(\frac{1}{3}+\frac{\cos\delta}{2\sqrt{3}} \ri)
\log_{2}\lf[\frac{1}{3}+\frac{\cos\delta}{2\sqrt{3}} \ri]
-\lf(\frac{2}{3}-\frac{\cos\delta}{2\sqrt{3}} \ri)
\log_{2}\lf[\frac{2}{3}-\frac{\cos\delta}{2\sqrt{3}} \ri]
\,,
\label{E3c}
\end{eqnarray}
where the superscript $(i,j;k)$ explicitly represents the specific composition
of the bipartitions $A_{2}=\{S_{i},S_{j}\}$ and $B_{1}=\{S_{k}\}$,
with $i,j,k=1,2,3$ and $i\neq j\neq k$.
Moreover, in order to simplify the notation, the definition
$E_{vN\,\alpha}^{(i,j;k)} \equiv E_{vN}^{(i,j;k)}(|W_{\alpha}^{(3)}(\delta)\ran)$
has been introduced.
The states $|W_{\alpha}^{(3)}(\delta)\ran$, with $\alpha=\mu,\tau$,
possess correlation properties dependent on $\delta$.

\begin{figure}[t]
\centering
\includegraphics*[width=8cm]{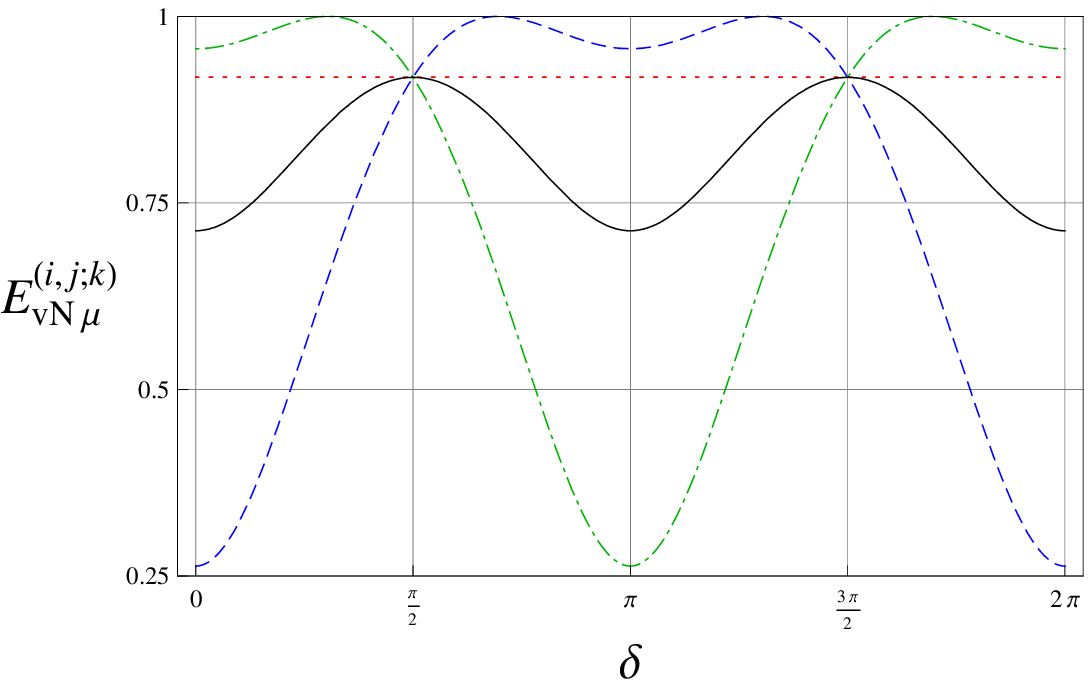}
\caption{(Color online) The von Neumann entropy $E_{vN\,\mu}^{(i,j;k)}$ and
the average von Neumann entropy $\lan E_{vN\,\mu}^{(2:1)} \ran$
as functions of the $CP$-violating phase $\delta$.
$E_{vN\,\mu}^{(i,j;k)}$ is plotted for the following bipartitions
$i,j;k$: $(a)$ $i=1$, $j=2$, and $k=3$ (dotted line);
$(b)$ $i=1$, $j=3$, and $k=2$ (dashed line);
$(c)$ $i=2$, $j=3$, and $k=1$ (dot-dashed line).
$E_{vN\,\mu}^{(1,2;3)}$ is constant and
takes the reference value $E_{21}^{(3)}\,=\,0.918296$.
The average entropy $\lan E_{vN\,\mu}^{(2:1)} \ran$ (full line)
attains the maximal value $E_{21}^{(3)}$
at $\delta=\frac{\pi}{2}\pm p \pi$, with $p$ integer.}
\label{FigW3_2}
\end{figure}

Let us, for instance, consider the state $|W_{\mu}^{(3)}(\delta)\ran$;
in Fig. \ref{FigW3_2}, the plots display the behavior of
$E_{vN\,\mu}^{(i,j;k)}$ and $\lan E_{vN\,\mu}^{(2:1)}\ran$
as a function of $\delta$ in the range $[0,2\pi]$.
While $E_{vN\,\mu}^{(1,2;3)}$ (dotted line) takes
the constant reference value $E_{21}^{(3)}$ (as for the state $|W^{(3)}\ran$),
the quantities $E_{vN\,\mu}^{(1,3;2)}$ (dashed line)
and $E_{vN\,\mu}^{(2,3;1)}$ (dot-dashed line)
vary with $\delta$, attaining the absolute maximum $1$ at the points
$\delta_{1} \,=\, \pm \arccos \lf( -\frac{1}{\sqrt{3}} \ri) \pm 2 p \pi$ and
$\delta_{2} \,=\, \pm \arccos \lf( \frac{1}{\sqrt{3}} \ri) \pm 2 p \pi$
(with $p$ integer), respectively.
Therefore, the state $|W_{\mu}^{(3)}(\delta_{i})\ran$, with $i=1,2$,
exhibits maximal entanglement in a given bipartition,
equal to the entanglement shown by the $GHZ$ state $|GHZ^{(3)}\ran$.
Moreover, for each given range of values of $\delta$, we see that
either $E_{vN\,\mu}^{(1,3;2)}$ (dashed line) or $E_{vN\,\mu}^{(2,3;1)}$
(dot-dashed line) exceeds the reference value $E_{21}^{(3)}$. This phenomenon of periodic
entanglement concentration is reminiscent of spin squeezing in collective
atomic variables. On the other hand, the average von Neumann entropy
$\lan E_{vN\,\mu}^{(2:1)} \ran$ stays below the reference value $E_{21}^{(3)}$,
attaining it at the points $\delta \,=\, \frac{\pi}{2} \pm p \pi$.
In conclusion, the free parameter $\delta$ can be used to concentrate
and squeeze the entanglement in a specific bipartition, allowing a sharply
peaked distribution of entanglement, corresponding to a lowering of the average
von Neumann entropy.

\subsection{Case of $|\underline{W}^{(4)}(\tilde{\theta}^{max};\tilde{\delta})\ran$ states}
\label{entW4}

\begin{figure}[ht]
\centering
\includegraphics*[width=8cm]{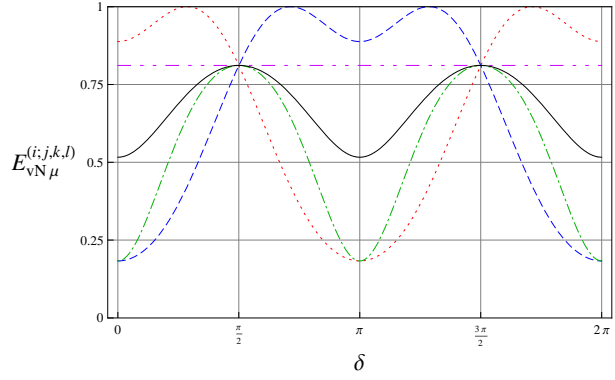}
\caption{(Color online) One-dimensional sections of the von Neumann entropies
$E_{vN\,\mu}^{(i;j,k,l)}$ for unbalanced $1:3$ bipartitions and
their average $\lan E_{vN\,\mu}^{(1:3)} \ran$ as functions of the phase
$\de$ ($\de \equiv \de_{14} = \de_{23}$). Similarly to the case of
three qubits, in the unbalanced four-qubit instance a concentration of
entanglement can be observed in the entropies
$E_{vN\,\mu}^{(1;2,3,4)}$ (dotted line) and $E_{vN\,\mu}^{(2;1,3,4)}$ (dashed line).
The entropy $E_{vN\,\mu}^{(4;1,2,3)}$ (double-dot-dashed line) is constant and
takes the reference value $E_{31}^{(4)} = 0.811278$.
The entropy $E_{vN\,\mu}^{(3;1,2,4)}$ (dot-dashed line),
and the average entropy $\lan E_{vN\,\mu}^{(1:3)} \ran$ (full line)
are always limited by this value, reaching it at points $\de = \frac{\pi}{2}+p\pi$.}
\label{FigW4_2sec}
\end{figure}

Due to the increased number of degrees of freedom, the class of
$W$-like states for $N=4$, i.e. Eq.~(\ref{fermix4}), yields a more
complex scenario for investigation. Proceeding as in Section \ref{entW3}, we
fix the rotation angles at their maximal values
$\theta_{ij}^{max}$, given by Eq.~(\ref{thetamax4}), and leave
free the phases $\delta_{ij}$. The matrix
$U^{(4f)}(\tilde{\theta};\tilde{\delta})$ acquires the form:
\begin{equation}
U^{(4f)}(\tilde{\delta}) \,=\,
\frac{1}{2}
\lf(
\ba{cccc} 1 & 1 & 1 & 1 \\
-1 -\frac{z_{14}}{3}-\frac{z_{23}^{*}}{3}
&1 -\frac{z_{14}}{3}-\frac{z_{23}^{*}}{3}
& -\frac{z_{14}}{3}+\frac{2z_{23}^{*}}{3}
& z_{14} \\
-\frac{1}{2}+\frac{z_{23}}{2}-\frac{z_{14}z_{34}^{*}}{3}
+\frac{z_{23}^{*}z_{34}^{*}}{6}+\frac{z_{34}^{*}}{2}
& -\frac{1}{2}-\frac{z_{23}}{2}-\frac{z_{14}z_{34}^{*}}{3}
+\frac{z_{23}^{*}z_{34}^{*}}{6}-\frac{z_{34}^{*}}{2}
& 1-\frac{z_{14}z_{34}^{*}}{3}-\frac{z_{23}^{*}z_{34}^{*}}{3}
& z_{34}^{*} z_{14} \\
\frac{1}{2}-\frac{z_{14}}{3}+ \frac{z_{23}^{*}}{6}
-\frac{z_{23}z_{34}}{2} +\frac{z_{34}}{2}
& -\frac{1}{2}-\frac{z_{14}}{3}+ \frac{z_{23}^{*}}{6}
+\frac{z_{23}z_{34}}{2} +\frac{z_{34}}{2}
& -\frac{z_{14}}{3}-\frac{z_{23}^{*}}{3}-z_{34}
& z_{14}
\ea
\ri)
\label{4fmatrixdelta}
\end{equation}
where $z_{ij}\,\equiv\,e^{i\delta_{ij}}$.
The explicit analytical expressions for the entanglement measures evaluated on the states
$|W_{\alpha}^{(4)}(\tilde{\delta})\ran \equiv |W_{\alpha}^{(4)}(\tilde{\theta}^{max};\tilde{\delta})\ran$
$(\alpha=e,\mu,\tau,s)$ are rather long and involved, and are reported in Appendix \ref{appendixW4}.
Note that the state $|W_{e}^{(4)}(\tilde{\delta})\ran$ coincides with the usual  $|W^{(4)}\ran$ state.
As an example, let us analyze in detail the entanglement
of the state $|W_{\mu}^{(4)}(\tilde{\delta})\ran$,
that depends on the phases $\de_{14}$ and $\de_{23}$
and is independent of the phase $\de_{13}$.
\begin{figure}[h]
\centering
\includegraphics*[width=15cm]{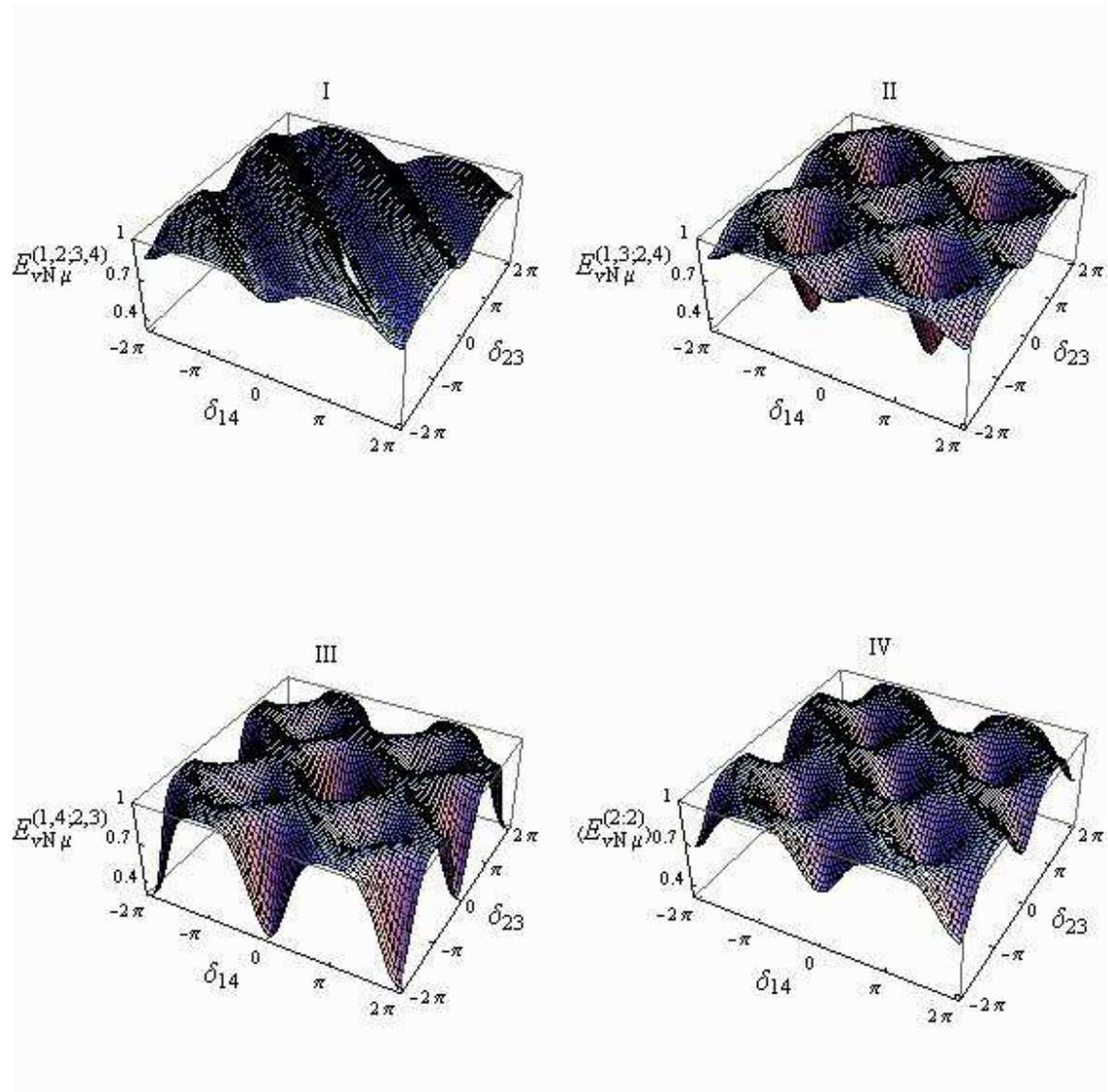}
\caption{(Color online) The von Neumann entropy $E_{vN\,\mu}^{(i,j;k,l)}$
for balanced bipartitions and the average von Neumann entropy
$\lan E_{vN\,\mu}^{(2:2)} \ran$ as functions of the phases $\de_{14}$ and $\de_{23}$.
Panel I shows $E_{vN\,\mu}^{(1,2;3,4)}$. It exhibits an oscillating behavior along the
direction parallel to the vector $(\de_{14}, \de_{23})=(1,1)$. Panels II and III show
the entropies $E_{vN\,\mu}^{(1,3;2,4)}$ and $E_{vN\,\mu}^{(1,4;2,3)}$, respectively.
They exhibit a nontrivial behavior yielding a periodic array structure of holes and dips.
The combined behaviors of all the entropies result in the average von Neumann entropy,
displayed in panel IV. All the four functions reach the maximum attainable value $1$ of
the entanglement at $\de_{14} + \de_{23} \,=\, \pm p \pi $, with $p$ odd integer.}
\label{FigW4_1}
\end{figure}
In Fig.~\ref{FigW4_1}, the plots I-III display
$E_{vN\,\mu}^{(1,2;3,4)}$, $E_{vN\,\mu}^{(1,3;2,4)}$, and
$E_{vN\,\mu}^{(1,4;2,3)}$, respectively, as a function of $\de_{14}$
and $\de_{23}$; the plot IV displays the behavior of the average
entropy $\lan E_{vN\,\mu}^{(2:2)} \ran$. The entanglement takes the
maximum value $1$ in correspondence of the values given in
Eq.~(\ref{deltamax4}), i.e. for $\de_{14} +\de_{23} \,=\, \pm p
\pi $, with $p$ odd integer. Moreover, while
$E_{vN\,\mu}^{(1,2;3,4)}$ exhibits an oscillating behavior along the
direction parallel to the vector $(\de_{14}, \de_{23})=(1,1)$, the
quantities $E_{vN\,\mu}^{(1,3;2,4)}$, $E_{vN\,\mu}^{(1,4;2,3)}$, and
$\lan E_{vN\,\mu}^{(2:2)} \ran$ show a periodic array structure of
holes. \\
Next, we consider the entropies corresponding to the
unbalanced bipartitions $E_{vN\,\mu}^{(i;j,k,l)}$.
The surface plots of these entropic measures,
as functions of $\de_{14}$ and $\de_{23}$,
are similar to those for the case of balanced bipartitions,
shown in Fig.~\ref{FigW4_1}.
In order to better highlight their structure, in Fig. \ref{FigW4_2sec}, we plot
one-dimensional sections of the surfaces belonging to the plane $\de_{14}=\de_{23}$.
We see that, as in the three qubit instance,
concentrations of entanglement (with a value in the range $[E_{31}^{(4)}\,,1]$)
occurs for the bipartitions $(1;2,3,4)$ and $(2;1,3,4)$, corresponding to
a lowering of the average entropy $\lan E_{vN\,2}^{(1:3)} \ran$.
In the range $[0,2\pi]$, both $E_{vN\,\mu}^{(1;2,3,4)}$ (dotted line)
and $E_{vN\,\mu}^{(2;1,3,4)}$ (dashed line) exceed in alternating order
the reference value $E_{31}^{(4)}$, and attain their maximum value $1$,
respectively at the points $\de_{a}=\pm \arccos \Big[\frac{3}{2}(\sqrt{2}-1)\Big]\pm 2p\pi$
and $\de_{b}=\pm \arccos \Big[-\frac{3}{2}(\sqrt{2}-1)\Big]\pm 2p\pi$. This behavior
is again reminiscent of spin squeezing in atomic systems.
Analogously to the three-qubit instance, the average entropy exhibits an oscillatory behavior,
and stays below the reference $E_{31}^{(4)}$, reaching it at $\de = \frac{\pi}{2}+p\pi$. \\
As they depend non trivially on all the phases $\de_{ij}$,
the states $|W_{\tau}^{(4)}(\tilde{\delta})\ran$ and
$|W_{s}^{(4)}(\tilde{\delta})\ran$ possess an even richer structure
of quantum correlations, compared to the case $|W_{\mu}^{(4)}(\tilde{\delta})\ran$
However, in both instances, one observes similar effects as the ones that occur
for the state  $|W_{\mu}^{(4)}(\tilde{\delta})\ran$.

\section{Quantifying entanglement in quark and neutrino flavor mixing}
\label{QuarksandNeutrino}

\begin{figure}[ht]
\centering
\includegraphics*[width=17cm]{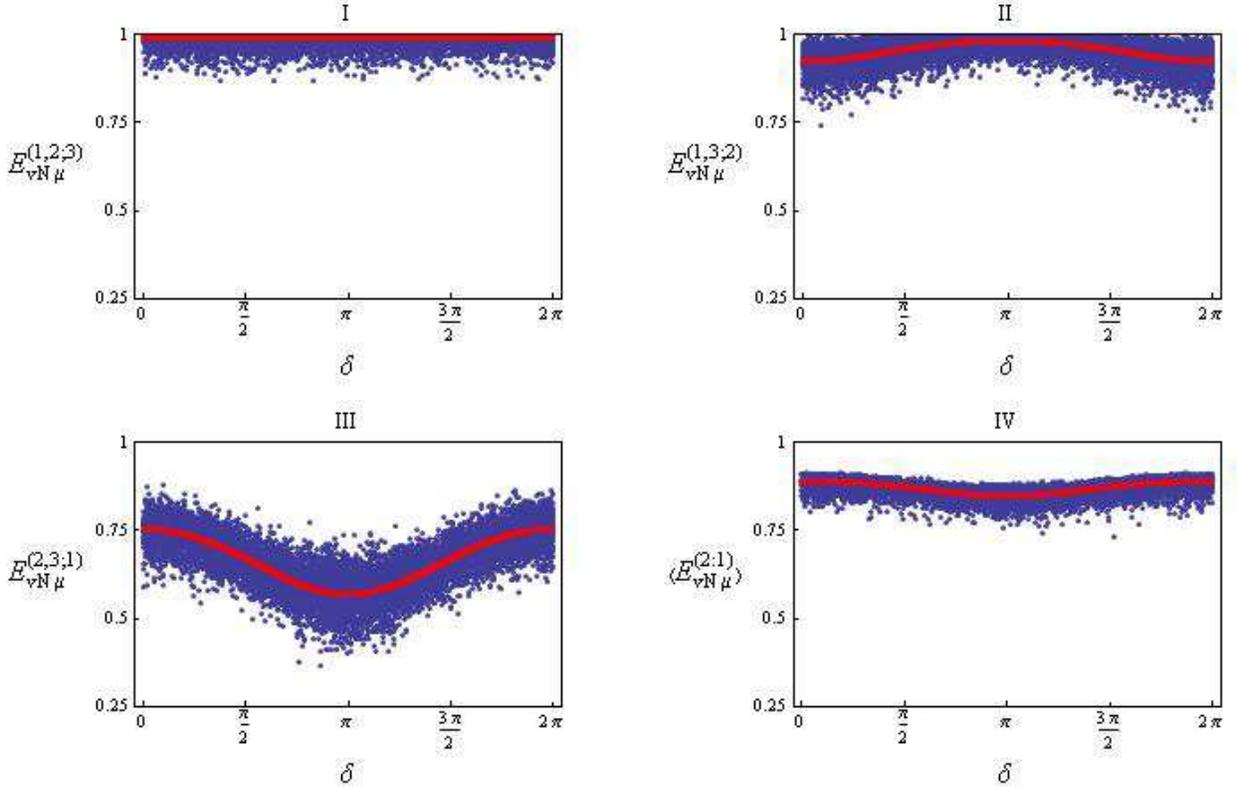}
\caption{(Color online) The von Neumann entropies $E_{vN\,\mu}^{(i,j;k)}$ for
all possible bipartitions and the average von Neumann entropy
$\lan E_{vN\,\mu}^{(2:1)} \ran$ as functions of the $CP$-violating
phase $\delta$. In panel I we plot the entropy $E_{vN\,\mu}^{(1,2;3)}$.
It is constant and close to $1$, the maximum attainable value of entanglement.
In panel II we plot the entropy $E_{vN\,\mu}^{(1,3;2)}$. It is moderately
$\de$-dependent and reaches its maximum at $\de=\pi$, still below $1$ (within
the experimental statistical errors). In panel III we plot the entropy $E_{vN\,\mu}^{(1,3;2)}$.
It corresponds to the bipartition with the least content of entanglement, is strongly
dependent on $\de$, and reaches a minimum at $\de=\pi$. The resulting average entropy
$\lan E_{vN\,\mu}^{(2:1)} \ran$, displayed in panel IV, is weakly $\de$-dependent and
reaches a minimum at $\de=\pi$. The mixing angles $\theta_{ij}^{MNSP}$ are assumed to be
Gaussian random variables, with a distribution centered at the mean values
$\overline{\theta}_{ij}^{MNSP}$ fixed to coincide with the experimental values
(\ref{Leptmixang}), and a standard deviation $\sigma_{ij}$ chosen to coincide
with $\frac{\delta\theta_{ij}^{MNSP}}{3}$. The uncertainties
$\delta\theta_{ij}^{MNSP}$ are fixed at the maximum values
between the left and right extrema given in Eq.~(\ref{Leptmixang}).
The thick full lines represent the entropies with $\theta_{ij}^{MNSP}=\overline{\theta}_{ij}^{MNSP}$,
and null uncertainty.}
\label{neutrent}
\end{figure}

In this Section, we quantify the entanglement in situations of
quarks or neutrino mixing, described by the three flavor states
defined in Eq.~(\ref{fermix3}). We will set the parameters
of the matrix (\ref{CKMmatrix}) at the values established by
the current experiments
\cite{ParticleData,Ohlsson,Maltoni,Fogli:2006yq}. In the case of
quarks, the mixing angles of the CKM matrix, are given by
\cite{Ohlsson}:
\begin{equation}
\theta_{12}^{CKM} \,=\, 13.0^{o}\pm 0.1^{o} \,, \quad
\theta_{13}^{CKM} \,=\, 0.2^{o}\pm 0.1^{o} \,, \quad
\theta_{23}^{CKM} \,=\, 2.4^{o}\pm 0.1^{o} \,.
\label{Quarkmixang}
\end{equation}
Moreover, a measurement of the $CP$ violation has yielded the value
for the $CP$-violating phase \cite{ParticleData}
\begin{equation}
\de^{CKM} \,=\, 1.05 \pm 0.24 \,.
\label{deltaCKM}
\end{equation}
In Table \ref{TabQuark}, we list the values for the von Neumann
entropies $E_{vN\,\alpha}^{(i,j;k)}$, with $\alpha = d',s',b'$ and
$i,j,k = d,s,b$, and $\langle E_{vN\,\alpha}^{(2:1)}\rangle$
corresponding to the states (\ref{fermix3}), with the mixing
angles and the $C$P-violating phase fixed to
Eqs.~(\ref{Quarkmixang}) and (\ref{deltaCKM}), respectively,
without taking into account the uncertainties.
\begin{table}[h]
\begin{tabular}{|c|c|c|c|c|}
\hline
$\alpha$ & $E_{vN\,\alpha}^{(d,s;b)}$ & $E_{vN\,\alpha}^{(d,b;s)}$ &
$E_{vN\,\alpha}^{(s,b;d)}$ & $\langle E_{vN\,\alpha}^{(2:1)}\rangle$
\\ \hline\hline
d' & $0.0002$ & $0.2889$ & $0.2890$ & $0.1927$ \\\hline
s' & $0.0185$ & $0.2960$ & $0.2887$ & $0.2011$ \\ \hline
b' & $0.0186$ & $0.0180$ & $0.0010$ & $0.0126$ \\ \hline
\end{tabular}
\caption{von Neumann entropies $E_{vN\,\alpha}^{(i,j;k)}$ and
$\langle E_{vN\,\alpha}^{(2:1)}\rangle$ $(\alpha = d',s',b')$ for
the three-flavor states associated with the quark mixing.}
\label{TabQuark}
\end{table}
We see that, in the range of the experimentally measured values of the mixing
angles, the entanglement stays low, very far from the maximum attainable value $1$.
Moreover, it concentrates in the bipartitions $(d,b;s)$ and $(s,b;d)$ of the
states $|d'\ran$ and $|s'\ran$, while it is very small for the state $|b'\ran$. \\
In the case of neutrinos, the most recent estimates of the
parameters of the MNSP matrix are expressed by the following
relations \cite{Fogli:2006yq}:
\begin{equation}
\sin^{2}\theta_{12}^{MNSP} \,=\,  0.314(1\begin{array}{c}
  +0.18 \\
  -0.15
\end{array} ) \,, \quad \sin^{2}\theta_{13}^{MNSP} \,=\,
(0.8  \begin{array}{c}
  +2.3 \\
  -0.8
\end{array} ) \times 10^{-2}\,, \quad \sin^{2}\theta_{23}^{MNSP} \,=\,  0.45
(1\begin{array}{c}
  +0.35 \\
  -0.20
\end{array} ) \,.
\label{Leptmixang}
\end{equation}
The $CP$-violating phase associated with lepton mixing is, at
present, completely undetermined; therefore, $\de^{MNSP}$ may take
an arbitrary value in the interval $[0,2\pi)$.
In Table \ref{TabNeutrino}, by using the relations
(\ref{Leptmixang}) (without taking into account the uncertainties)
and for arbitrary $\de^{MNSP}$, we list the entropies corresponding
to the neutrino flavor states. The given intervals of possible values
are obviously due to the freedom in the choice of the $CP$-violating phase.
\begin{table}[h]
\begin{tabular}{|c|c|c|c|c|}
\hline
$\alpha$ & $E_{vN\,\alpha}^{(1,2;3)}$ & $E_{vN\,\alpha}^{(1,3;2)}$ &
$E_{vN\,\alpha}^{(2,3;1)}$ & $\langle E_{vN\,\alpha}^{(2:1)}\rangle$
  \\ \hline\hline
  $e$ & $0.0672$ & $0.8948$ & $0.9038$ & $0.5995$ \\\hline
  $\mu$ & $0.9916$ & $0.9220 - 0.9813$ & $0.5679 - 0.7536$ & $0.8469 - 0.8891$ \\ \hline
  $\tau$ & $0.9939$ & $0.8397 - 0.9352$ & $0.4784 - 0.6922$ & $0.8025 - 0.8419$ \\ \hline
\end{tabular}
\caption{von Neumann entropies $E_{vN\,\alpha}^{(i,j;k)}$ and
$\langle E_{vN\,\alpha}^{(2:1)}\rangle$ $(\alpha=e,\mu,\tau)$ for the
three-flavor states associated with the neutrino mixing.}
\label{TabNeutrino}
\end{table}
Comparing Tables \ref{TabQuark} and \ref{TabNeutrino}, it turns out
that the neutrino mixing states are more entangled and their entanglement
is more homogeneously distributed among the different bipartions, compared
to the quark mixing states.
In the case of neutrinos, the uncertainties are very large.
Moreover, the value taken by the mixing angle $\theta_{13}^{MNSP}$
is crucial. In fact, only if such an angle is non-vanishing, then
the entropies are dependent on the $CP$-violating phase. Therefore,
it is interesting to investigate the behavior of entanglement when
one takes into account the experimental uncertainties on the mixing angles.
To this aim, we assume that $\theta_{ij}^{MNSP}$ takes random values
normally distributed around the experimentally observed values.
For instance, in Fig.~\ref{neutrent}, we plot
$E_{vN\,\mu}^{(i,j;k)}$ and $\lan E_{vN\,\mu}^{(2:1)} \ran$ as a
function of the free parameter $\delta^{MNSP}\equiv\delta$.
We see that the entanglement corresponding to bipartitions
$(1,2;3)$ and $(1,3;2)$ keeps high, (panels I and II); on the other side, the
bipartition $(2,3;1)$ exhibits lower amount of entanglement (panel
III), leading to a lowering of the average amount of global entanglement (panel IV).
Thus, we can conclude that, for the states $| \nu_{\mu} \rangle$,
the parties $2$ and $3$ are more strongly correlated compared to the pairs $1,2$, and
$1,3$. Similar conclusions hold for the states $| \nu_{e} \rangle$ and
$| \nu_{\tau} \rangle$.

\section{Decoherence in neutrino oscillations}
\label{DecoStatBeam}

In the previous Sections, we have provided an analysis of the mixing effect
in terms of the quantum correlations of multipartite mode-entangled states,
by exploiting tools commonly used in the domain of quantum information theory.
The characterization of the entanglement of generalized multipartite $W$ states,
through the measurement of the amount of the quantum information content
of these states, constitutes a description of a fundamental effect
of particle physics.
The physical insight of such an analysis acquires even more relevance if
it is transferred to a dynamical scenario, by studying the phenomenon of
particle oscillations.
Let us recall that both the phenomenon of particle oscillations and quantum entanglement
are due to the superposition principle which gives place to coherent interference
among the different mass eigenstates.
In the particular instance of neutrinos, the standard theory of oscillations
is developed in the plane-wave approximation \cite{neutroscillplanewave}.
Adopting such an approximation, all the results obtained in the previous Sections
hold for any time in the free evolution dynamics.
However, a more realistic description of the phenomenon
can be achieved by means of the wave packet approach
\cite{Nussinov,GiuntiKim,Giunti2,Giunti:2008cf},
for reviews see Refs.~\cite{Beuthe,Giunti:2007ry}.
The three massive neutrinos possess different masses;
consequently, the corresponding wave packets propagate at different speeds,
and acquire an increasing spatial separation with respect each other.
Therefore, the free evolution leads to a natural lowering of the coherent
interference effects, associated with the destruction of the oscillation phenomenon
and with the vanishing of the multipartite quantum entanglement.
In this Section, we intend to analyze the quantum correlations of multipartite
entangled neutrino states by using the wave packet description for massive neutrinos.
In particular, we want to study the ``decoherence'' effects,
induced by the free evolution, on the multipartite entanglement among neutrino mass
eigenstates.
Let us notice that the forthcoming analysis, as well all the formalism developed
in this work, can be applied to any system exhibiting the particle mixing.

\begin{figure}[h]
\centering
\includegraphics*[width=17cm]{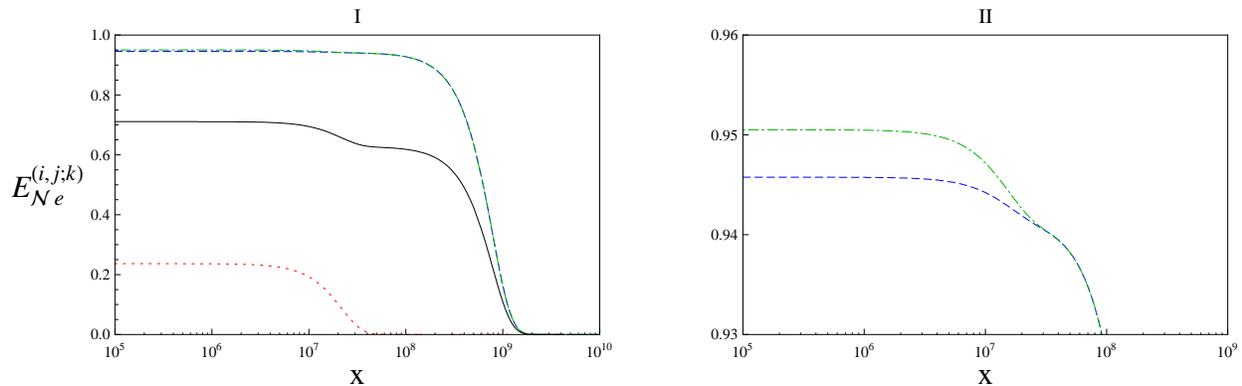}
\caption{(Color online) The logarithmic negativities $E_{\mathcal{N}\,e}^{(i,j;k)}$
for all possible bipartitions and the average logarithmic negativity
$\lan E_{\mathcal{N}\,e}^{(2:1)} \ran$ as functions of the distance $x$.
The quantities $E_{\mathcal{N}\,e}^{(1,3;2)}$ (dashed line)
and $E_{\mathcal{N}\,e}^{(2,3;1)}$ (dot-dashed line), see panel I,
show a high amount of entanglement content in the corresponding bipartitions,
and seem to be superimposed.
In panel II we plot a zoom of $E_{\mathcal{N}\,e}^{(1,3;2)}$ and
$E_{\mathcal{N}\,e}^{(2,3;1)}$ to observe the differences in their behaviors:
the two curves are initially separated, and then they superimpose each other.
The bipartition $(1,2;3)$, associated with the quantity $E_{\mathcal{N}\,e}^{(1,2;3)}$
(dotted line), exhibits the lowest amount of entanglement.
The full line corresponds to the average logarithmic negativity
$\lan E_{\mathcal{N}\,e}^{(2:1)} \ran$.
The mixing angles $\theta_{ij}^{MNSP}$ and the squared mass differences $\Delta m_{ij}^{2}$
are fixed at the experimental values (\ref{Leptmixang}) and (\ref{sqmassdiff}), respectively.
We assume the values $E = 10 \,GeV$, $\sigma_{p} = 1 \,GeV$,
and $\xi=0$ for the remaining parameters in Eq.~(\ref{statwavepack}).
All the plotted quantities are independent of the CP-violating phase $\delta$,
that can be assumed arbitrary.
The $x$ axis is in logarithmic scale, and the dimensions are meters.
}
\label{deconeutrente}
\end{figure}

Following the procedure developed in Refs.~\cite{GiuntiKim,Giunti2,Giunti:2008cf},
by considering one only one space dimension,
a neutrino with definite flavor, propagating along the $x$ direction.
can be described by the state:
\begin{equation}
|\nu_{\alpha}(x,t)\rangle \,=\, \sum_{j} U_{\alpha,j} \, \psi_{j}(x,t) \,|\nu_{j}\rangle \,,
\label{neutwvpack}
\end{equation}
where the $U_{\alpha,j}$ denotes the corresponding element of the mixing matrix,
$|\nu_{j}\rangle$ is the mass eigenstate with mass $m_{j}$,
and $\psi_{j}(x,t)$ is its wave function. Assuming for the momentum of the
massive neutrino $|\nu_{j}\rangle$ a Gaussian distribution $\psi_{j}(p)$,
the wave function is given by:
\begin{equation}
\psi_{j}(x,t) \,=\, \frac{1}{\sqrt{2\pi}} \,
\int \,dp \, \psi_{j}(p) \, e^{i p x -i E_{j}(p) t} \,, \quad
\psi_{j}(p) \,=\, \frac{1}{(2\pi \sigma_{p}^{2})^{1/4}} \,
e^{-\frac{1}{4\sigma_{p}^{2}}(p-p_{j})^{2}} \,,
\label{wavfunc}
\end{equation}
where $p_{j}$ is the average momentum, $\sigma_{p}$ is
the momentum uncertainty, and $E_{j}(p) \,=\, \sqrt{p^{2}+m_{j}^{2}}$.
The density matrix associated with the pure state Eq.~(\ref{neutwvpack}) writes:
\begin{equation}
\rho_{\alpha}(x,t) \,=\, |\nu_{\alpha}(x,t)\rangle \langle \nu_{\alpha}(x,t)| \,.
\label{densmatwvpack}
\end{equation}
If the inequality $\sigma_{p}\ll E_{j}^{2}(p_{j})/m_{j}$ holds,
the energy $E_{j}(p)$ can be approximated by $E_{j}(p)\simeq E_{j}+v_{j}(p-p_{j})$,
with $E_{j}\equiv \sqrt{p_{j}^{2}+m_{j}^{2}}$, and $v_{j}\equiv
\frac{\partial E_{j}(p)}{\partial p}\big|_{p=p_{j}} \,=\,\frac{p_{j}}{E_{j}} $
is the group velocity of the wave packet
of the massive neutrino $|\nu_{j}\rangle$.
In this case, the integration over $p$ in Eq.~(\ref{wavfunc}) is Gaussian
and can be easily performed, yielding the following expression for $\rho_{\alpha}(x,t)$
\begin{equation}
\rho_{\alpha}(x,t) \,=\, \frac{1}{\sqrt{2\pi\sigma_{x}^{2}}} \sum_{j,k} U_{\alpha j} U_{\alpha k}^{*}
e^{-i (E_{j}-E_{k})t+i(p_{j}-p_{k})x-\frac{1}{4\sigma_{x}^{2}}[(x-v_{j}t)^{2}+(x-v_{k}t)^{2}] }
|\nu_{j}\rangle \langle\nu_{k}| \,,
\label{wavepack2}
\end{equation}
where $\sigma_{x} \,=\, (2\sigma_{p})^{-1}$.
In the instance of extremely relativistic neutrinos,
the following approximations are usually assumed
\begin{equation}
E_{j} \,\simeq \, E + \xi \frac{m_{j}^{2}}{2E} \,, \qquad
p_{j} \,\simeq \, E-(1-\xi)\frac{m_{j}^{2}}{2E} \,, \qquad
v_{j} \,\simeq \, 1-\frac{m_{j}^{2}}{2E_{j}^{2}}
\label{relapprox}
\end{equation}
where $E$ is the neutrino energy in the limit of zero mass,
and $\xi$ is a dimensionless constant depending on the characteristic
of the production process \cite{GiuntiKim,Giunti2}.
The density matrix (\ref{wavepack2}) provides a space-time description
of neutrino dynamics.
However, in realistic situations, it is convenient to consider
the corresponding stationary process, which is associated with the
time-independent density matrix $\rho_{\alpha}(x)$
obtained by the time average of $\rho_{\alpha}(x,t)$  \cite{Giunti2}.
By taking into account Eq.~(\ref{relapprox}), and by computing a Gaussian
integration over the time, the density matrix becomes  \cite{Giunti2}
\begin{equation}
\rho_{\alpha}(x) \,=\, \sum_{j,k} U_{\alpha j} U_{\alpha k}^{*}
\exp\left[-i \frac{\Delta m_{jk}^{2} x}{2E}
-\left( \frac{\Delta m_{jk}^{2} x}{4\sqrt{2}E^{2}\sigma_{x}}\right)^{2}
- \left(\xi \frac{\Delta m_{jk}^{2}}{4\sqrt{2}E \sigma_{p}}\right)^{2} \right]
|\nu_{j}\rangle \langle\nu_{k}| \,,
\label{statwavepack}
\end{equation}
with $\Delta m_{jk}^{2} \,=\, m_{j}^{2}-m_{k}^{2}$.
The density matrix (\ref{statwavepack}) can be used to study, in the wave packet approach,
the phenomenon of neutrino oscillations for stationary neutrino beams
\cite{GiuntiKim,Giunti2,Giunti:2008cf}. \\
Here, we intend to analyze the coherence of the quantum superposition
of the neutrino mass eigenstates,
by looking at the spatial behavior of the multipartite entanglement
of the state (\ref{statwavepack}).
By establishing the identification
$|\nu_{i}\rangle \,=\, |\delta_{i,1}\rangle_{1}|\delta_{i,2}\rangle_{2}|\delta_{i,3}\rangle_{3}
\equiv |\delta_{i,1}\delta_{i,2}\delta_{i,3}\rangle$ $(i=1,2,3)$,
we can easily construct from Eq.~(\ref{statwavepack}) the matrix with elements
$\langle lmn|\rho_{\alpha}(x)|ijk \rangle$, where $i,j,k,l,m,n \,=\, 0,1$.
Let us notice that the density matrix $\rho_{\alpha}(x)$ describes a mixed state,
whose non-diagonal elements are suppressed by a Gaussian function of $x$.
An appropriate quantifier of multipartite entanglement for the state $\rho_{\alpha}(x)$
is based on the set of logarithmic negativities defined in subsection \ref{averlogNegativity}.
We analytically compute the quantities $E_{\mathcal{N}\,\alpha}^{(i,j;k)}$,
for $i,j,k=1,2,3$ and $i\neq j\neq k$, and the average logarithmic negativity
$\langle E_{\mathcal{N}\,\alpha}^{(2:1)} \rangle$, for the neutrino states
with flavor $\alpha \,=\, e, \mu, \tau$.
We assume for the mixing angles $\theta_{ij}^{MNSP}$ the experimental values
(\ref{Leptmixang}).
The squared mass differences are fixed at the experimental values reported in
Ref.~\cite{Fogli:2006yq}:
\begin{eqnarray}
&&\Delta m_{21}^{2} \,=\, \delta m^{2} \,, \qquad
\Delta m_{31}^{2} \,=\, \Delta m^{2} + \frac{\delta m^{2}}{2} \,, \qquad
\Delta m_{32}^{2} \,=\, \Delta m^{2} - \frac{\delta m^{2}}{2} \,, \nonumber  \\
&& \delta m^{2} \,=\, 7.92 \times 10^{-5} \, eV^{2} \,, \qquad\quad
\delta m^{2} \,=\, 2.6 \times 10^{-3} \, eV^{2} \,.
\label{sqmassdiff}
\end{eqnarray}
The parameters $E$ and $\sigma_{p}$ in Eq.~(\ref{statwavepack})
are fixed at the values $E = 10 \,GeV$ and $\sigma_{p} = 1 \,GeV$.
Moreover, although depending on the particular production process \cite{Giuntixi},
the parameter $\xi$ is put to zero for simplicity.
In Fig.~\ref{deconeutrente}, we plot the logarithmic negativities for the
electronic neutrino, i.e. $E_{\mathcal{N}\,e}^{(i,j;k)}$ as function of the
distance $x$.
The bipartitions $(1,3;2)$ and $(2,3;1)$, see panel I, exhibit a high entanglement
content $(>0.93)$ that keeps almost constant for $x \lesssim 10^{8}\, m$;
finally, it goes to zero for $x \approx 3 \times 10^{9}\, m$.
The bipartition $(1,2;3)$ exhibits a low entanglement $(<0.24)$,
that goes to zero for $x \approx 9 \times 10^{7}\, m$.
Furthermore, let us remark that the the logarithmic negativities
$E_{\mathcal{N}\,e}^{(i,j;k)}$ and $\lan E_{\mathcal{N}\,e}^{(2:1)} \ran$
for the electronic neutrino are independent of the CP-violating phase $\delta$.

\begin{figure}[h]
\centering
\includegraphics*[width=17cm]{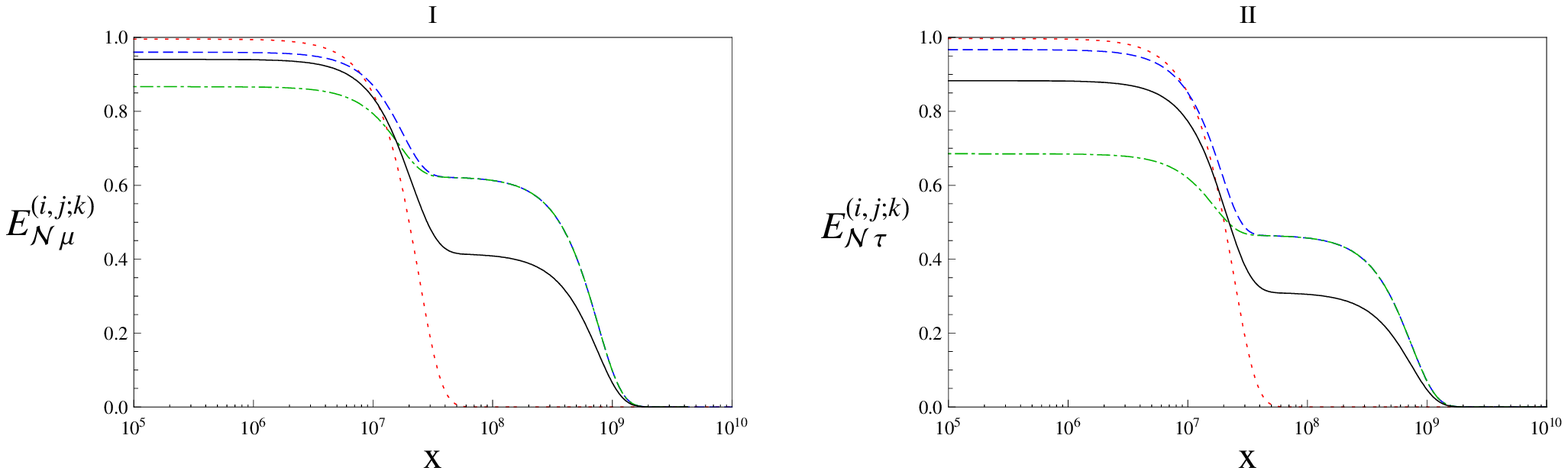}
\caption{(Color online) The logarithmic negativities $E_{\mathcal{N}\,\alpha}^{(i,j;k)}$
for all possible bipartitions and the average logarithmic negativity
$\lan E_{\mathcal{N}\,\alpha}^{(2:1)} \ran$, with $\alpha=\mu,\,\tau$,
as functions of the distance $x$.
In panel I we plot the negativities for the muonic neutrino.
The bipartition $(1,2;3)$, associated with the quantity
$E_{\mathcal{N}\,\mu}^{(1,2;3)}$ (dotted line),
shows the highest initial amount of entanglement,
that goes to zero for a lower of $x$ with respect to the other bipartitions.
$E_{\mathcal{N}\,\mu}^{(1,3;2)}$ (dashed line) and
$E_{\mathcal{N}\,\mu}^{(2,3;1)}$ (dot-dashed line)
show peculiar behaviors, that consist in alternating
slowly decreasing and rapidly decreasing slopes.
The average logarithmic negativity $\lan E_{\mathcal{N}\,\mu}^{(2:1)} \ran$
(full line) summarizes the behavior of the global entanglement.
In panel II we plot the negativities for the tauonic neutrino;
The behaviors of the negativies for the tauonic instance are similar
to the negativities for the muonic instance.
The curves associated to a given bipartition are plotted
with the same plotstyle.
The mixing angles $\theta_{ij}^{MNSP}$ and the squared mass differences $\Delta m_{ij}^{2}$
are fixed at the experimental values (\ref{Leptmixang}) and (\ref{sqmassdiff}), respectively.
We assume the values $E = 10 \,GeV$, $\sigma_{p} = 1 \,GeV$,
and $\xi=0$ for the remaining parameters in Eq.~(\ref{statwavepack}).
The CP-violating phase $\delta$ is put to zero.
The $x$ axis is in logarithmic scale, and the dimensions are meters.
}
\label{deconeutrentmutau}
\end{figure}

In the muonic and tauonic instances, the independence from the CP-violating phase
$\delta$ holds no more.
Therefore, first we choose to study the quantum correlations of these states
for $\delta = 0$; then we consider separately the influence of a non-zero $\delta$.
In Fig.~\ref{deconeutrentmutau}, we plot the logarithmic negativities for the
muonic and tauonic neutrinos as functions of the distance $x$ with $\delta =0$.
We see that the spatial behavior of multipartite entanglement for muonic
and tauonic neutrinos are similar.
The logarithmic negativities
$E_{\mathcal{N}\,\mu}^{(1,2;3)}$ and $E_{\mathcal{N}\,\tau}^{(1,2;3)}$
are initially close to $1$, and they go to zero for $x \approx 10^{8} \, m$.
On the other side, $E_{\mathcal{N}\,\mu}^{(1,3;2)}$, $E_{\mathcal{N}\,\mu}^{(2,3;1)}$,
$E_{\mathcal{N}\,\tau}^{(1,3;2)}$, and $E_{\mathcal{N}\,\tau}^{(2,3;1)}$
exhibit alternating regimes with slowly decreasing slope
and with rapidly decreasing slope;
moreover, all vanish for $x \approx  3\times 10^{9} \, m$.

The average logarithmic negativity
$\lan E_{\mathcal{N}\,\alpha}^{(2:1)} \ran$
can be used to define a {\it decoherence length} $L_{decoh}$ as
\begin{equation}
L_{decoh} \, : \, \lan E_{\mathcal{N}\,\alpha}^{(2:1)} \ran\, (L_{decoh}) \,=\, 0 \,.
\label{Ldecoh}
\end{equation}
From Figs.~\ref{deconeutrente}, \ref{deconeutrentmutau},
for assigned experimental parameters, we see that the common decoherence length
for the neutrinos of flavor $\alpha = e,\mu,\tau$ can be estimated at a value of
$L_{decoh} \approx 3 \times 10^{6} Km$.

\begin{figure}[h]
\centering
\includegraphics*[width=17cm]{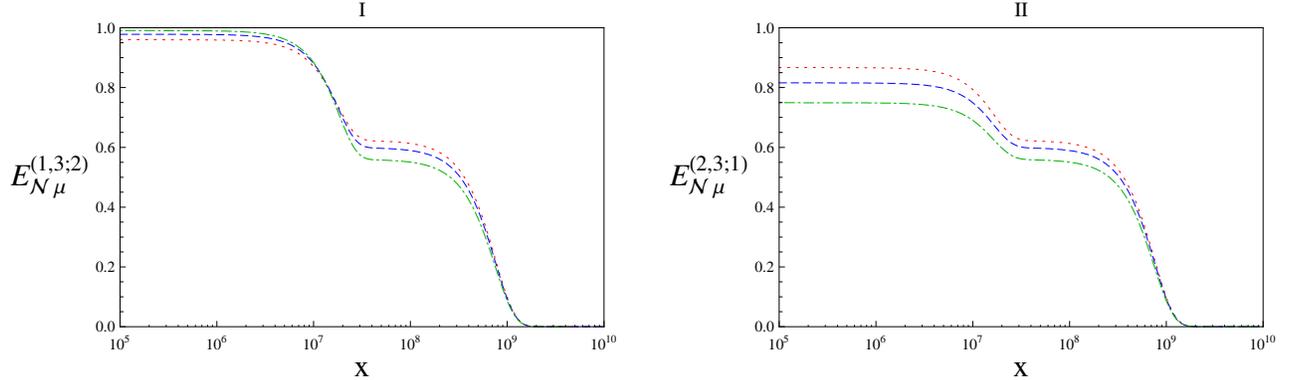}
\caption{(Color online) The logarithmic negativities
$E_{\mathcal{N}\,\mu}^{(1,3;2)}$ (panel I)
and $E_{\mathcal{N}\,\mu}^{(2,3;1)}$ (panel II) as functions of the distance $x$
for different choices of the CP-violating phase $\delta$: (a) $\delta = 0$ (dotted line);
(b) $\delta = \frac{\pi}{2}$ (dashed line); (b) $\delta = \pi$ (dot-dashed line).
$E_{\mathcal{N}\,\mu}^{(1,2;3)}$ is independent of $\delta$.
The mixing angles $\theta_{ij}^{MNSP}$, the squared mass differences $\Delta m_{ij}^{2}$,
the parameters $E$, $\sigma_{p}$, and $\xi$ are fixed as in Figs.~\ref{deconeutrente} and \ref{deconeutrentmutau}.
The $x$ axis is in logarithmic scale, and the dimensions are meters.
}
\label{deconeutrentmuDeCP}
\end{figure}

Finally, we consider the influence of a non-vanishing phase $\delta$
in determining the spatial behavior of multipartite entanglement of
stationary neutrino beams.
To this aim, in Fig.~\ref{deconeutrentmuDeCP} we plot the logarithmic negativities
for the muonic neutrino
$E_{\mathcal{N}\,\mu}^{(1,3;2)}$ and $E_{\mathcal{N}\,\mu}^{(2,3;1)}$,
with $\delta$ fixed at the values $\delta = 0, \frac{\pi}{2} , \pi$.
The behavior of $E_{\mathcal{N}\,\mu}^{(1,2;3)}$ is not reported as
it is independent of $\delta$.
We observe that the CP-violating phase $\delta$ does not lead to a
change of the decoherence length $L_{decoh}$.
However, we see that it may lead a lowering or an increasing of the amount of
entanglement in a given bipartition, in agreement with the results obtained
for the instance of static neutrinos.
Similar results can be obtained for the tauonic instance.

\section{Conclusions}

The study of entanglement between field modes can be fruitfully applied to a large variety
of quantum mechanical systems, either in the usual case of many-particle multipartite entangled
states or in the more intriguing instance of single-particle multipartite entangled states. In
the present paper, stimulated by recent work on single-particle nonlocality and entanglement in
quantum optical systems, we have extended the analysis of mode entanglement to systems of
elementary particle physics. In particular, we have determined and studied the structural properties
of the multipartite entangled states that occur in the physics of flavor mixing, either in quark or
in leptonic systems. These states are generalizations of the well known $W$ states, endowed with nontrivial
relative phases. These states include, as a special instance, the symmetric $W$ state and the set of
states orthogonal to it. We have implemented global and statistical approaches, based on the
distribution of different bipartite entanglements, to quantify the generic aspects of multipartite
entanglement in such states. We have studied in detail the correlation properties of three- and four-flavor
$W$ states. For properly chosen mixing parameters, we have shown that the phases, responsible for the
$CP$-violation effects in particle physics, can be used to concentrate the entanglement in a particular
bipartition, and we have identified some periodic patterns of entanglement concentration, dispersion, and
revivals, that are reminiscent of spin-squeezing phenomena for the collective variables of many-body atomic
systems. Moreover, we have analyzed the entanglement for the three-quark and three-neutrino mixing. In the
particular instance of neutrino mixing, we have determined the effects of the free relative phases on the
distribution of entanglement. By exploiting the wave packet treatment for neutrino mass eigenstates,
we have considered in detail the influence of decoherence induced by the free evolution
on the multipartite entanglement. A decoherence length can be defined as the distance associated with
vanishing average global entanglement. Finally, we have studied the role of the CP-violating phase
in the dynamics of free propagation.

\section{Acknowledgments}

We acknowledge financial support from MIUR, under PRIN 2005 National
Research Project, from INFN, and from INFM-CNR Coherentia Research
and Development Center. F. I. acknowledges financial support from
ISI Foundation.

\appendix

\section{Entropic measures for the states $|W_{q}^{(4)}(\tilde{\delta})\ran$}
\label{appendixW4}

Below we report the analytical expressions for the eigenvalues
corresponding to the reduced density matrices of the states
$|W_{\alpha}^{(4)}(\tilde{\delta})\ran$ $(\alpha=e,\mu,\tau,s)$.
Let us denote by $\ul{\la}_{\alpha}^{(i;j,k,l)}$ and $\ul{\la}_{\alpha}^{(i,j;k,l)}$
the eigenvalue vectors associated with the reduced density matrices
$Tr_{j,k,l}[|W_{\alpha}^{(4)}(\tilde{\delta})\ran\lan W_{\alpha}^{(4)}(\tilde{\delta})|]$
and $Tr_{k,l}[|W_{\alpha}^{(4)}(\tilde{\delta})\ran\lan W_{\alpha}^{(4)}(\tilde{\delta})|]$,
respectively.
We get
\begin{eqnarray}
\ul{\la}_{e}^{(1;2,3,4)} \,=\,&&
\ul{\la}_{e}^{(2;1,3,4)} \,=\,
\ul{\la}_{e}^{(3;1,2,4)} \,=\,
\ul{\la}_{e}^{(4;1,2,3)} \,=\,
\ul{\la}_{\mu}^{(4;1,2,3)} \,=\,
\ul{\la}_{\tau}^{(4;1,2,3)} \,=\,
\ul{\la}_{s}^{(4;1,2,3)} \,=\,
\frac{1}{4}\{3\,,1\}
\,,
\label{E4a} \\
&& \nonumber \\
\ul{\la}_{\mu}^{(1;2,3,4)} \,=\,&&
\frac{1}{36}\Big\{25-6\cos\de_{14}-6\cos\de_{23}-2\cos(\de_{14}+\de_{23})\,,
11+6\cos\de_{14}+6\cos\de_{23}+2\cos(\de_{14}+\de_{23})\Big\}
\,,
\label{E4b}
\\
&& \nonumber \\
\ul{\la}_{\mu}^{(2;1,3,4)} \,=\,&&
\frac{1}{36}\Big\{11-6\cos\de_{14}-6\cos\de_{23}+2\cos(\de_{14}+\de_{23})\,,
25+6\cos\de_{14}+6\cos\de_{23}-2\cos(\de_{14}+\de_{23})\Big\}
\,,
\label{E4c}
\\
&& \nonumber \\
\ul{\la}_{\mu}^{(3;1,2,4)} \,=\,&&
\frac{1}{36}\Big\{5-4\cos(\de_{14}+\de_{23})\,,31+4\cos(\de_{14}+\de_{23})\Big\}
\,,
\label{E4d}
\end{eqnarray}
\begin{eqnarray}
\ul{\la}_{\tau}^{(1;2,3,4)} \,=\,&&
\frac{1}{72}\Big\{16-6\cos\de_{14}-6\cos\de_{23}-2\cos(\de_{14}+\de_{23})
+6\cos(\de_{14}-\de_{34})-6\cos(\de_{14}-\de_{23}-\de_{34})
\nonumber \\
&&-9\cos\de_{34}
+6\cos(\de_{23}+\de_{34})+3\cos(2\de_{23}+\de_{34})\,,
56+6\cos\de_{14}+6\cos\de_{23}+2\cos(\de_{14}+\de_{23})\nonumber \\
&&-6\cos(\de_{14}-\de_{34})+
6\cos(\de_{14}-\de_{23}-\de_{34})+9\cos\de_{34}-6\cos(\de_{23}+\de_{34})-3\cos(2\de_{23}+\de_{34})\Big\}
\,,
\label{E4e}
\\
&& \nonumber \\
\ul{\la}_{\tau}^{(2;1,3,4)} \,=\,&&
\frac{1}{72}\Big\{56-6\cos\de_{14}-6\cos\de_{23}+2\cos(\de_{14}+\de_{23})
-6\cos(\de_{14}-\de_{34})-6\cos(\de_{14}-\de_{23}-\de_{34})
\nonumber \\
&&-9\cos\de_{34}
-6\cos(\de_{23}+\de_{34})+3\cos(2\de_{23}+\de_{34})\,,
16+6\cos\de_{14}+6\cos\de_{23}-2\cos(\de_{14}+\de_{23})\nonumber \\
&&+6\cos(\de_{14}-\de_{34})+
6\cos(\de_{14}-\de_{23}-\de_{34})+9\cos\de_{34}+6\cos(\de_{23}+\de_{34})-3\cos(2\de_{23}+\de_{34})\Big\}
\,,
\label{E4f}
\\
&& \nonumber \\
\ul{\la}_{\tau}^{(3;1,2,4)} \,=\,&&
\frac{1}{36}\Big\{11+2\cos(\de_{14}+\de_{23})-6\cos(\de_{14}-\de_{34})-6\cos(\de_{23}+\de_{34})\,,
\nonumber \\
&&25-2\cos(\de_{14}+\de_{23})+6\cos(\de_{14}-\de_{34})+6\cos(\de_{23}+\de_{34})\Big\}
\,,
\label{E4g}
\end{eqnarray}
\begin{eqnarray}
\ul{\la}_{s}^{(1;2,3,4)} \,=\,&&
\frac{1}{72}\Big\{56+6\cos\de_{14}+6\cos\de_{23}+2\cos(\de_{14}+\de_{23})
+6\cos(\de_{14}-\de_{34})-6\cos(\de_{14}-\de_{23}-\de_{34})
\nonumber \\
&&-9\cos\de_{34}
-6\cos(\de_{23}+\de_{34})+3\cos(2\de_{23}+\de_{34})\,,
16-6\cos\de_{14}-6\cos\de_{23}-2\cos(\de_{14}+\de_{23})\nonumber \\
&&-6\cos(\de_{14}-\de_{34})+
6\cos(\de_{14}-\de_{23}-\de_{34})+9\cos\de_{34}-6\cos(\de_{23}+\de_{34})-3\cos(2\de_{23}+\de_{34})
\Big\}
\,,
\label{E4h}
\\
&& \nonumber \\
\ul{\la}_{s}^{(2;1,3,4)} \,=\,&&
\frac{1}{72}\Big\{16+6\cos\de_{14}+6\cos\de_{23}-2\cos(\de_{14}+\de_{23})-6\cos(\de_{14}-\de_{34})
-6\cos(\de_{14}-\de_{23}-\de_{34})
\nonumber \\
&&-9\cos\de_{34}-6\cos(\de_{23}+\de_{34})+3\cos(2\de_{23}+\de_{34}) \,,
56-6\cos\de_{14}-6\cos\de_{23}+2\cos(\de_{14}+\de_{23})
\nonumber \\
&&+6\cos(\de_{14}-\de_{34})
+6\cos(\de_{14}-\de_{23}-\de_{34})
+9\cos\de_{34}+6\cos(\de_{23}+\de_{34})-3\cos(2\de_{23}+\de_{34})\Big\}
\,,
\label{E4i}
\\
&& \nonumber \\
\ul{\la}_{s}^{(3;1,2,4)} \,=\,&&
\frac{1}{36}
\Big\{25-2\cos(\de_{14}+\de_{23})-6\cos(\de_{14}-\de_{34})-6\cos(\de_{23}+\de_{34}) \,,
\nonumber \\
&& 11+2\cos(\de_{14}+\de_{23})+6\cos(\de_{14}-\de_{34})+6\cos(\de_{23}+\de_{34})
\Big\} \, ,
\label{E4l}
\end{eqnarray}
\begin{eqnarray}
\ul{\la}_{e}^{(1,2;3,4)} \,=\,&&
\ul{\la}_{e}^{(1,3;2,4)} \,=\,
\ul{\la}_{e}^{(1,4;2,3)} \,=\,
\frac{1}{2}\{0\,,0\,,1\,,1\} \, ,
\label{E4I} \\
&& \nonumber \\
\ul{\la}_{\mu}^{(1,2;3,4)} \,=\,&&
\frac{1}{18}\Big\{0\,,0\,,7-2\cos(\de_{14}+\de_{23})\,,
11+2\cos(\de_{14}+\de_{23})\Big\} \, ,
\label{E4II}
\\
&& \nonumber \\
\ul{\la}_{\mu}^{(1,3;2,4)} \,=\,&&
\frac{1}{18}\Big\{0\,,0\,,10-3\cos\de_{14}-3\cos\de_{23}+\cos(\de_{14}+\de_{23})\,,
8+3\cos\de_{14}+3\cos\de_{23}-\cos(\de_{14}+\de_{23})\Big\} \, ,
\label{E4III}
\\
&& \nonumber \\
\ul{\la}_{\mu}^{(1,4;2,3)} \,=\,&&
\frac{1}{18}\Big\{0\,,0\,,8-3\cos\de_{14}-3\cos\de_{23}-\cos(\de_{14}+\de_{23})\,,
10+3\cos\de_{14}+3\cos\de_{23}+\cos(\de_{14}+\de_{23})\Big\} \, ,
\label{E4IV}
\end{eqnarray}
\begin{eqnarray}
\ul{\la}_{\tau}^{(1,2;3,4)} \,=\,&&
\frac{1}{18}\Big\{0\,,0\,,10+\cos(\de_{14}+\de_{23})-3\cos(\de_{14}-\de_{34})-3\cos(\de_{23}+\de_{34})\,,
8-\cos(\de_{14}+\de_{23})\nonumber \\
&&+3\cos(\de_{14}-\de_{34})+3\cos(\de_{23}+\de_{34})\Big\} \, ,
\label{E4V}
\\
&& \nonumber \\
\ul{\la}_{\tau}^{(1,3;2,4)} \, = \, &&
\frac{1}{72}\Big\{0\,,0\,,38-6\cos\de_{14}-6\cos\de_{23}+2\cos(\de_{14}+\de_{23})
+6\cos(\de_{14}-\de_{34}) -6\cos(\de_{14}-\de_{23}-\de_{34})\nonumber \\
&&-9\cos\de_{34}-6\cos(\de_{23}+\de_{34})+3\cos(2\de_{23}+\de_{34})\,,
34+6\cos\de_{14}+6\cos\de_{23}-2\cos(\de_{14}+\de_{23})\nonumber \\
&&+6\cos(\de_{14}-\de_{34})+6\cos(\de_{14}-\de_{23}-\de_{34})+9\cos\de_{34}
+6\cos(\de_{23}+\de_{34})-3\cos(2\de_{23}+\de_{34})\Big\} \, ,
\label{E4VI}
\\
&& \nonumber \\
\ul{\la}_{\tau}^{(1,4;2,3)} \,=\,&&
\frac{1}{72}\Big\{0\,,0\,,34-6\cos\de_{14}-6\cos\de_{23}-2\cos(\de_{14}+\de_{23})
+6\cos(\de_{14}-\de_{34}) -6\cos(\de_{14}-\de_{23}-\de_{34})\nonumber \\
&&-9\cos\de_{34}+6\cos(\de_{23}+\de_{34})+3\cos(2\de_{23}+\de_{34})\,,
38+6\cos\de_{14}+6\cos\de_{23}+2\cos(\de_{14}+\de_{23})\nonumber \\
&&-6\cos(\de_{14}-\de_{34})+6\cos(\de_{14}-\de_{23}-\de_{34})+9\cos\de_{34}-6\cos(\de_{23}+\de_{34})
-3\cos(2\de_{23}+\de_{34})\Big\},
\label{E4VII}
\end{eqnarray}
\begin{eqnarray}
\ul{\la}_{s}^{(1,2;3,4)} \,=\,&&
\frac{1}{18}\Big\{0\,,0\,,8-\cos(\de_{14}+\de_{23})-3\cos(\de_{14}-\de_{34})-3\cos(\de_{23}+\de_{34})\,,
10+\cos(\de_{14}+\de_{23})\nonumber \\
&&+3\cos(\de_{14}-\de_{34})+3\cos(\de_{23}+\de_{34})\Big\} \, ,
\label{E4VIII}
\\
&& \nonumber \\
\ul{\la}_{s}^{(1,3;2,4)} \,=\,&&
\frac{1}{72}\Big\{0\,,0\,,34+6\cos\de_{14}+6\cos\de_{23}-2\cos(\de_{14}+\de_{23})
-6\cos(\de_{14}-\de_{34}) -6\cos(\de_{14}-\de_{23}-\de_{34})\nonumber \\
&&-9\cos\de_{34}-6\cos(\de_{23}+\de_{34})+3\cos(2\de_{23}+\de_{34})\,,
38-6\cos\de_{14}-6\cos\de_{23}+2\cos(\de_{14}+\de_{23})\nonumber \\
&&+6\cos(\de_{14}-\de_{34})+6\cos(\de_{14}-\de_{23}-\de_{34})+9\cos\de_{34}
+6\cos(\de_{23}+\de_{34})-3\cos(2\de_{23}+\de_{34})\Big\},
\label{E4IX}
\\
&& \nonumber \\
\ul{\la}_{s}^{(1,4;2,3)} \,=\,&&
\frac{1}{72}\Big\{0\,,0\,,38+6\cos\de_{14}+6\cos\de_{23}+2\cos(\de_{14}+\de_{23})
+6\cos(\de_{14}-\de_{34}) -6\cos(\de_{14}-\de_{23}-\de_{34})\nonumber \\
&&-9\cos\de_{34}+6\cos(\de_{23}+\de_{34})+3\cos(2\de_{23}+\de_{34})\,,
34-6\cos\de_{14}-6\cos\de_{23}-2\cos(\de_{14}+\de_{23})\nonumber \\
&&-6\cos(\de_{14}-\de_{34})+6\cos(\de_{14}-\de_{23}-\de_{34})+9\cos\de_{34}-6\cos(\de_{23}+\de_{34})
-3\cos(2\de_{23}+\de_{34})\Big\}.
\label{E4X}
\end{eqnarray}

The von Neumann entropies can be easily written as

\begin{equation}
E_{vN\,\alpha}^{(\cdot)} \,=\, -\sum_{n}\la_{\alpha}^{(\cdot)}(n)\log_{2}\la_{\alpha}^{(\cdot)}(n) \,.
\end{equation}

\end{document}